\documentclass[preprint,12pt]{elsarticle}

\usepackage{amssymb}
\usepackage{amsmath}
\usepackage[utf8]{inputenc}
\usepackage{graphicx}
\usepackage{multirow}
\usepackage{subcaption}
\usepackage{url}
\usepackage{pdflscape}
\usepackage{float}
\usepackage{bm}

\renewcommand{\vec}[1]{\bm{#1}}

\journal{}

\begin{document}

\begin{frontmatter}


\author{Tristan Millington\corref{cor1}}
\ead{t.millington@soton.ac.uk}
\author{Mahesan Niranjan}
\cortext[cor1]{Corresponding Author}

\title{Construction of Minimum Spanning Trees from Financial Returns using Rank Correlation}



\address{Department of Electronics and Computer Science, University of Southampton, Southampton, SO17 1BJ, UK}

\begin{abstract}
    The construction of minimum spanning trees (MSTs) from correlation matrices is an often used method to study relationships in the financial markets. However most of the work on this topic tends to use the Pearson correlation coefficient, which relies on the assumption of normality and can be brittle to the presence of outliers, neither of which is ideal for the study of financial returns. In this paper we study the inference of MSTs from daily US, UK and German financial returns using Pearson and two rank correlation methods, Spearman and Kendall's $\tau$. MSTs constructed using these rank methods tend to be more stable and maintain more edges over the dataset than those constructed using Pearson correlation. The edge agreement between the Pearson and rank MSTs varies significantly depending on the state of the markets, but the rank MSTs generally show strong agreement at all times. Deviation from univariate normality can be related to changes in the correlation matrices but is more difficult to connect to changes in the MSTs.  Irrelevant of coefficient, the trees tend to have similar topologies. Portfolios constructed from the MST correlation matrices have a smaller turnover than those from the full covariance matrix for the larger markets, but not for the smaller German market. Using a bootstrap method we find that the correlation matrices constructed using the rank correlations are more robust, but there is little difference between the robustness of the MSTs. 
\end{abstract}



\begin{keyword}
networks \sep correlation \sep finance \sep minimum spanning trees


\end{keyword}

\end{frontmatter}



\section{Introduction}

Investors tend to not invest purely in single asset, but due to a desire to reduce risk and increase diversification, own portfolios made up of multiple assets. To accurately assess risk in these portfolios we must understand the dynamics of the relationships between said assets.
Various methods of inferring the strength of relationships exist, for instance correlation \cite{mantegna1999} \cite{onnela2004clustering}, partial correlation \cite{millington_2019} \cite{kennet2010} \cite{millington_2020} \cite{wang2018correlation} or mutual information \cite{fiedor2014information} \cite{guo_2018}, but Pearson correlation is the most ubiquitous. From these asset to asset relationships we can use network theory to study the system as a whole.

Accurate inference of these correlation matrices from a dataset with $p$ assets and $n$ samples is challenging for numerous reasons. If $n$ is not significantly larger than $p$, the correlation matrix will contain significant amounts of noise. Due to non-stationarities in the markets, we often want a small window of data, where we can assume that the data is stationary \cite{tsay2005analysis}. This often results in $n$ being close to $p$, even if we have a large amount of data. Options to help remove noise in this situation include topological filtration methods (e.g. Minimum Spanning Trees \cite{mantegna1999}, Planar Maximally Filtered Graphs \cite{tumminello201040}), random matrix theory approaches \cite{namaki20113835} \cite{laloux2000random} \cite{plerou2002random} or thresholding \cite{boginski2005431} \cite{onnela2004clustering}.
In this paper we focus on the use of topological methods for this purpose, due to the simplicity of construction and interpretation. To start with, we follow Mantegna \cite{mantegna1999} and construct a distance matrix ($D$) from the correlation matrix ($C$):
\begin{equation}
    \label{eq:dist}
    D_{ij} = \sqrt{2 (1 - C_{ij})}
\end{equation}

This distance matrix is then used as the adjacency matrix for a new graph, the distance graph. From this we can use Kruskal's algorithm to create a minimum spanning tree (MST), which proceeds as follows
\begin{itemize}
	\item Initialise the tree as a disconnected graph made up of the nodes in the distance graph
	\item Sort edges of the distance graph in ascending order and place in a list
	\item For each edge between $i$ and $j$ in this sorted list
	\begin{itemize}
		\item If $i$ and $j$ are not in the same component in the tree, add this edge into the tree
	\end{itemize}
\end{itemize}
Various `stylised facts' are known about these trees, for instance 
\begin{itemize}
    \item Branches of the trees tend to contain companies in the same sector \cite{mantegna1999}
    \item The trees shrink and have different structures during times of market stress compared to market calm \cite{onnela2003dynamics} \cite{zhang20112020}
    \item The trees tend to have a `scale free' structure, with nodes of high degree (hubs) occurring more than would be expected from a random graph \cite{bonanno_2003} \cite{vandewalle2001non} \cite{onnela2003dynamics}
    \item Assets with large weights in Markowitz portfolios tend to be peripheral \cite{onnela2003247} \cite{huttner2018portfolio}
    \cite{peralta2016157}, however there is disagreement over whether to chose assets on the peripheries or center of the networks for better Sharpe ratios \cite{kaya2013eccentricity} \cite{pozzi2013spread} \cite{peralta2016157}
    \item The MSTs tend to only keep significant correlations \cite{aste_2010}
\end{itemize}

While most of the focus has been on the US markets, MST based models have been applied to markets from other countries (e.g. Japan \cite{jung2008537}, the UK \cite{coelho2007615}, Italy \cite{coletti2016246} and South Korea \cite{jung2006263}), to cryptocurrencies \cite{stosic2018collective} \cite{song2019121339} and to networks from neuroscience \cite{tewarie2014308} \cite{tewarie2015177}.

While the interpretability of the Pearson correlation coefficient is a big plus, it assumes normality, something which most assets return distributions do not follow \cite{cont_2001}, and is sensitive to outliers. There are of course correlation measures that do not suffer from these issues, namely those based on rank. Rank correlation methods calculate the correlation between the ranks of variables, which tends to remove the effects of outliers while still giving a measure of the degree to which two variables increase or decrease together. 

Most of the literature which studies the correlations between asset returns tends to use the Pearson correlation coefficient and so in this paper we compare networks inferred from stock returns using Pearson, Spearman and Kendall's $\tau$ correlation in order to see if the robustness of these rank correlations can improve our understanding of the stock markets.

Previous work \cite{micciche_2003} has briefly mentioned that MSTs constructed using Spearman correlation from volatility measures of stocks are more robust, but they did not explicitly compare the two correlation coefficients. In a paper more broadly looking at the effects of weighting observations, Pozzi et al. \cite{pozzi_2012} compare Pearson correlation and Kendall's $\tau$. They find that matrices constructed using Kendall's $\tau$ tend to contain more information than those constructed using Pearson correlation, and are affected less when they weight observations. 
A paper on a more similar theme to this is written by Musmeci et al. \cite{musmeci2017multiplex}, who take a multilayer network approach. Each layer is composed of a Planar Maximally Filtered Graph, constructed using a different method. Four methods are used to quantify relationships between assets, Pearson correlation, Kendall's $\tau$, tail dependence and partial correlation. They find that these layers tend to have significant differences, with between 30\% and 70\% of the edges being unique to each layer. Pearson and Kendall's $\tau$ tend to be the methods that agree the most, with a correlation of around 0.7 on the degree of nodes. Interestingly they find that the level of agreement drops during times of crisis, showing that these different methods tend to pick up different signals from the markets and indicating that being mindful of multiple methods of quantifying relationships is valuable when taking a network approach to financial returns.
The final example we found is by Shirokikh et al. \cite{shirokikh_2013} who use a thresholding model with Spearman correlation, but they do not compare how this model differs to a Pearson based one. 

To the best of our knowledge however, there has been no work which compares Pearson and Spearman correlation, or compares different rank correlations. Furthermore we are unaware of any work that compares Pearson and rank correlation MSTs. Therefore we undertake a detailed study on these fronts. 

The Pearson correlation between two variables $r_i$ and $r_j$ is defined as follows
	\begin{equation}
	C_{ij} = \frac{\sum_{i=1}^n (r_i(t) - \bar{r_i}) (r_j(t) - \bar{r_j})}{\sqrt{\sum_{i=1}^n ((r_i(t) - \bar{r_i})^2 (r_j(t) - \bar{r_j})^2 )}}
	\end{equation}

To calculate the Spearman correlation, we firstly sort the values, replace each value with its rank, and calculate the Pearson correlation between the ranks. This then measures the degree to which two variables monotonically increase or decrease together. Kendall's $\tau$ is slightly more complicated, measuring the relationship by considering the number of concordant pairs vs the number of discordant pairs. A pair of observations $(r_i(t), r_j(t)), (r_i(t+1), r_j(t+1))$ is concordant if $r_i(t) > r_j(t)$ and $r_i(t+1) > r_j(t+1)$ or if $r_i(t) < r_j(t)$ and $r_i(t+1) < r_j(t+1)$. It is discordant if $r_i(t) > r_j(t)$ and $r_i(t+1) < r_j(t+1)$ or if $r_i(t) < r_j(t)$ and $r_i(t+1) > r_j(t+1)$. The $\tau$-a formula simply counts the number of concordant pairs vs the number of discordant pairs, divided by the total number of pairs. This does not however take into account any ties that might occur in the data, so we use the $\tau$-b formulation, defined as
\begin{equation}
    \tau = \frac{n_c - n_d}{\sqrt{(n_0 - n_1)(n_0 - n_2)}}
\end{equation}
where $n_c$ is the number of concordant pairs, $n_d$ is the number of discordant pairs, $n_0 = \frac{n(n-1)}{2}$, $n_1 = \sum t_a (t_a - 1) / 2$, $n_2 = \sum u_a (u_a - 1) / 2$, $t_a$ is the number of values in the $a$th group of ties for variable $i$, $u_a$ is the number of values in the $a$th group of ties for variable $j$. Any reference to $\tau$ in the rest of the paper refers to this $\tau$-b formation.

\section{Data and Software}

	The data we use is downloaded from Yahoo Finance. For the UK data we use the FTSE100 companies, for the US data we use the S\&P500 companies and for the German data we use the DAX30 companies. We then use log returns from 2000/03/01 to 2019/10/21. For each dataset any company missing more than 10\% of its data is removed, and any missing values are filled forwards. If the values are missing from the start we backfill from the first good value. From this data we take a window of 504 days and slide along 30 days at a time. This leaves us with 4789 days of returns data from 229 companies for the US, 5063 days of returns data from 70 companies for the UK and 5066 days for 23 companies for Germany. Germany will also be referred to as DE in the rest of this paper. 
	Each company is tagged with a sector from the GICS classification using information from Bloomberg. This places each company into 1 of 11 sectors, Technology, Real Estate, Materials, Communications, Energy, Financials, Utilities, Industrials, Consumer Discretionary, Healthcare or Consumer Staples. 

We make use of Python, NumPy and SciPy \cite{oliphant2006guide} for general scripting, pandas \cite{mckinney-proc-scipy-2010} for handing the data, statsmodels \cite{seabold2010statsmodels} for some of the statistical analysis, matplotlib \cite{matplotlib} for plotting, arch \cite{kevin_sheppard_2019_2613877} for the implementation of the circular bootstrap, TopCorr for the MST construction (\url{https://github.com/shazzzm/topcorr}) and Networkx \cite{scipyproceedings_11} for the network analysis. The code and data is available at \url{https://github.com/shazzzm/rank_correlation_msts}.

\section{Results and Analysis}

\subsection{Correlation Matrix Analysis}

\begin{figure}
    \centering
    \begin{subfigure}{.30\textwidth}
          \includegraphics[width=\textwidth]{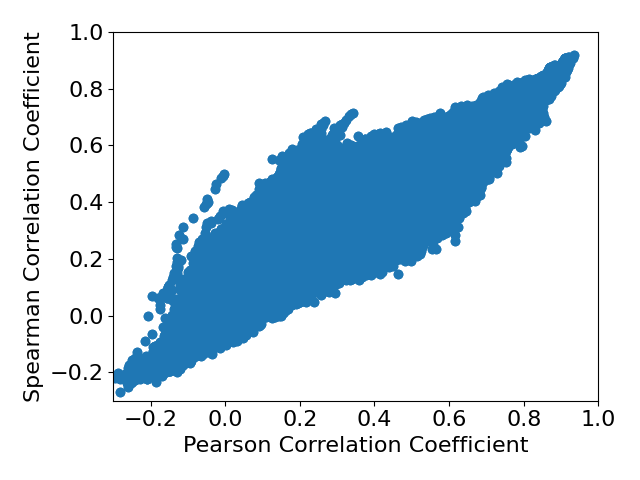}
          \caption{Pearson vs Spearman - US}
        \label{fig:pearson_vs_spearman_us}
    \end{subfigure}
    \begin{subfigure}{.30\textwidth}
          \includegraphics[width=\textwidth]{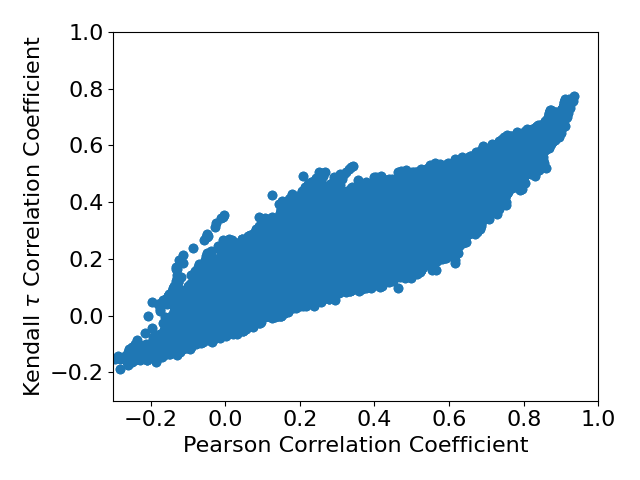}
          \caption{Pearson vs $\tau$ - US}
    \label{fig:pearson_vs_tau_us}
    \end{subfigure}
    \begin{subfigure}{.30\textwidth}
          \includegraphics[width=\textwidth]{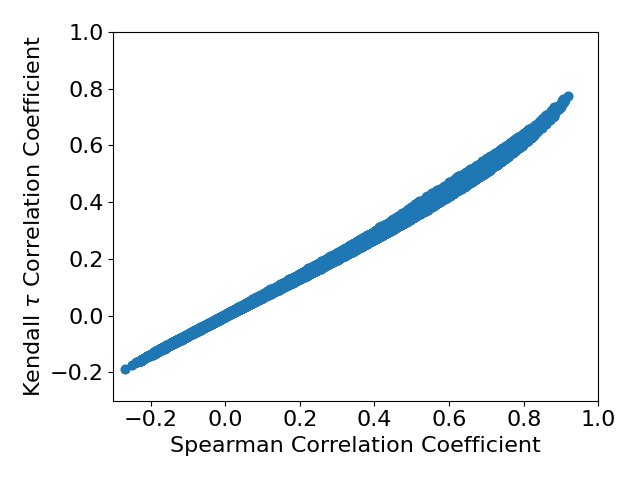}
          \caption{Spearman vs $\tau$ - US}
        \label{fig:spearman_vs_tau_us}
    \end{subfigure}
        \begin{subfigure}{.30\textwidth}
          \includegraphics[width=\textwidth]{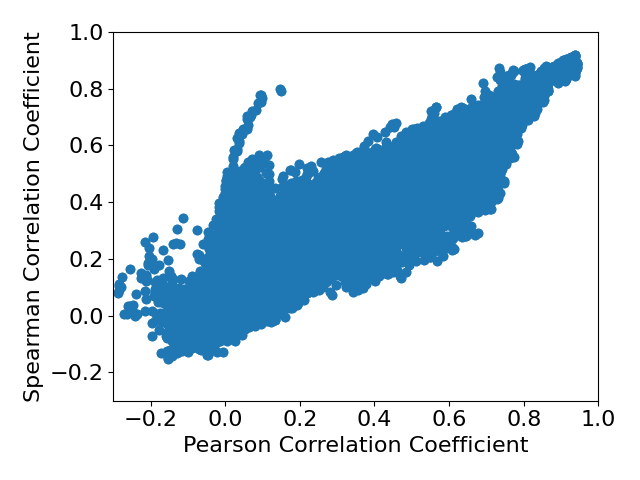}
          \caption{Pearson vs Spearman - UK}
        \label{fig:pearson_vs_spearman_uk}
    \end{subfigure}
    \begin{subfigure}{.30\textwidth}
          \includegraphics[width=\textwidth]{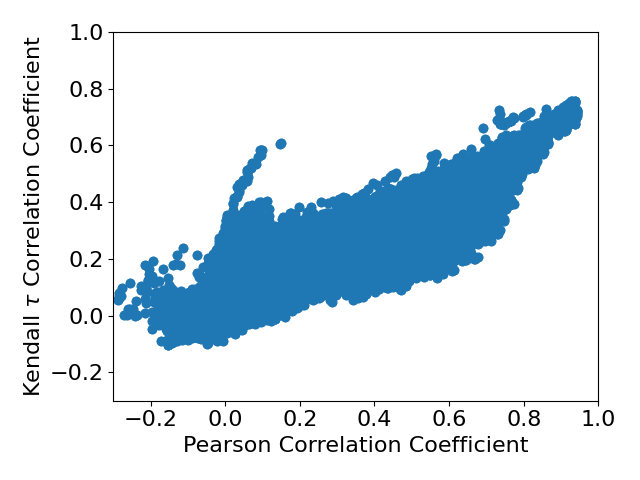}
          \caption{Pearson vs $\tau$ - UK}
    \label{fig:pearson_vs_tau_uk}
    \end{subfigure}
    \begin{subfigure}{.30\textwidth}
          \includegraphics[width=\textwidth]{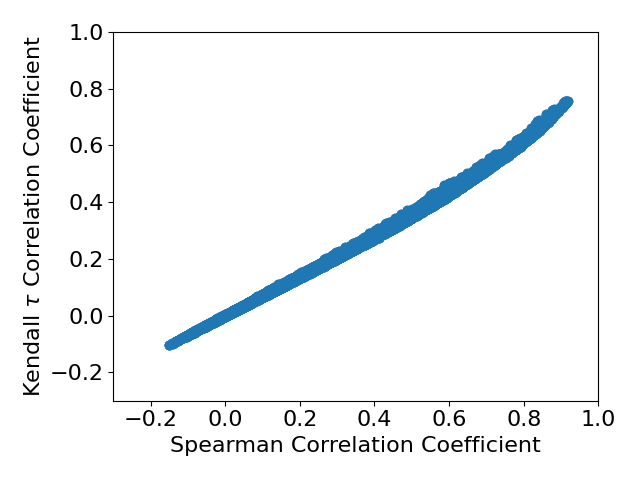}
          \caption{Spearman vs $\tau$ - UK}
        \label{fig:spearman_vs_tau_uk}
    \end{subfigure}
    \begin{subfigure}{.30\textwidth}
          \includegraphics[width=\textwidth]{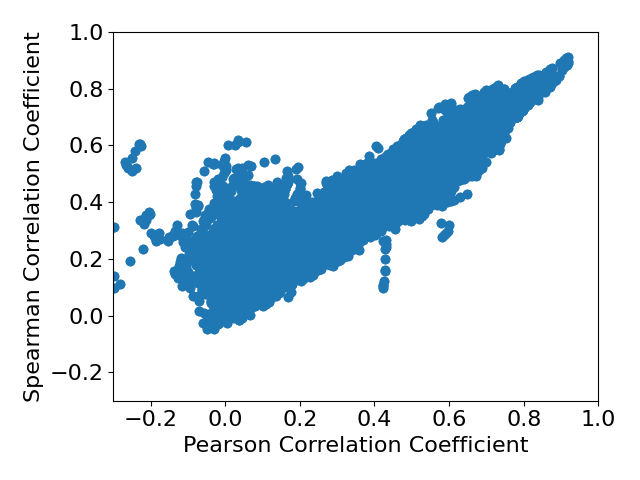}
          \caption{Pearson vs Spearman- DE}
        \label{fig:pearson_vs_spearman_de}
    \end{subfigure}
    \begin{subfigure}{.30\textwidth}
          \includegraphics[width=\textwidth]{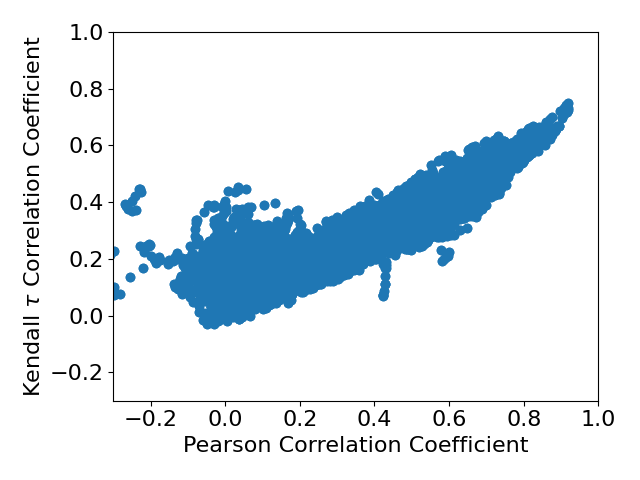}
          \caption{Pearson vs $\tau$ - DE}
    \label{fig:pearson_vs_tau_de}
    \end{subfigure}
    \begin{subfigure}{.30\textwidth}
          \includegraphics[width=\textwidth]{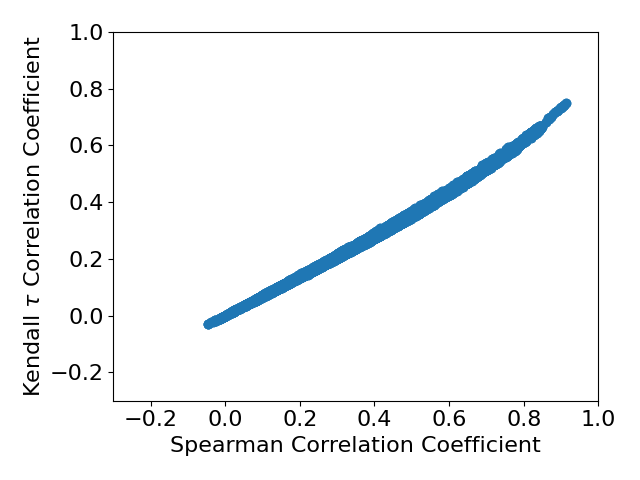}
          \caption{Spearman vs $\tau$ - DE}
        \label{fig:spearman_vs_tau_de}
    \end{subfigure}
    \caption{Relationship of the correlation coefficients for each country across the entirety of each dataset. There is a large degree of agreement and larger correlations are more likely to be similar, but the `fat' middle is notable when comparing the Pearson and rank correlations. The rank correlations themselves are relatively similar, with the $\tau$ correlations being smaller than the Spearman correlations.}
    \label{fig:correlation_comparison}
\end{figure}

Firstly we analyze the full correlation matrices with no filtration. A starting point is to look at the correlation coefficient for the same set of values. Figure \ref{fig:correlation_comparison} shows a set of scatter plots comparing the correlations. From this we can see there is a degree of agreement between all, and generally larger correlations are more likely to be similar. However there is a `fat' middle when comparing the rank correlations to the Pearson correlation, where there can be significant disagreement. Spearman and $\tau$ seem to be very similar, with there being a strong relationship between the two, although the $\tau$ correlations are slightly smaller than the Spearman ones. 

\begin{figure}
    \centering
    \begin{subfigure}{.30\textwidth}
          \includegraphics[width=\textwidth]{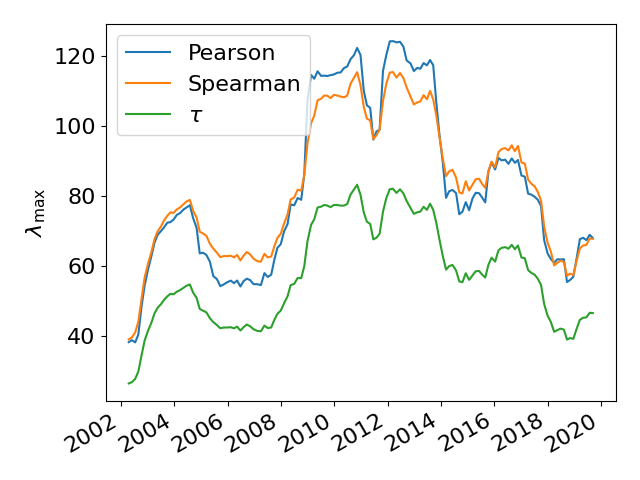}
          \caption{US}
        \label{fig:max_eig_us}
    \end{subfigure}
    \begin{subfigure}{.30\textwidth}
          \includegraphics[width=\textwidth]{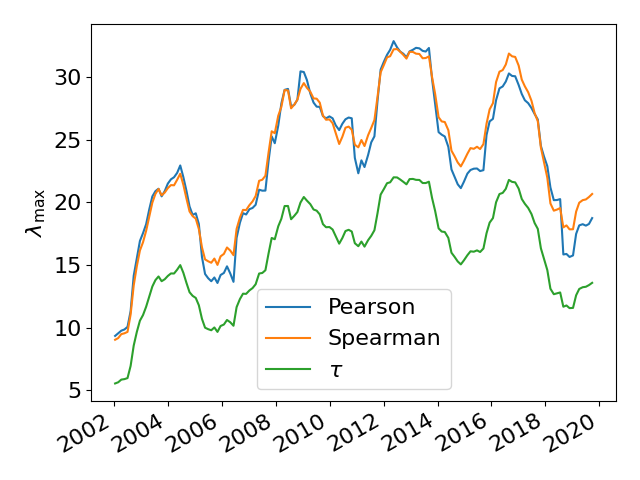}
          \caption{UK}
    \label{fig:max_eig_uk}
    \end{subfigure}
    \begin{subfigure}{.30\textwidth}
          \includegraphics[width=\textwidth]{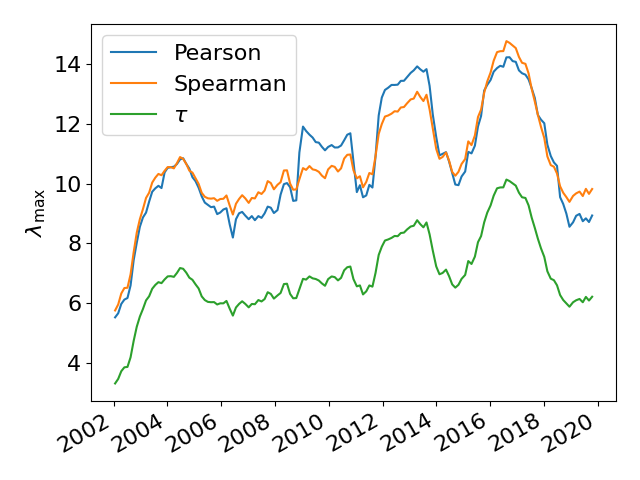}
          \caption{DE}
        \label{fig:max_eig_de}
    \end{subfigure}    
    \caption{Largest eigenvalue ($\lambda_{\max}$) in the networks over time. From this we can see the Spearman correlation has a slightly smaller largest eigenvalue than the Pearson correlation, while the $\tau$ correlation is much smaller. The rank methods also have a smaller range than the Pearson correlation. The volatility of the markets at times of stress is likely to lead to more outliers, so the robustness of the rank correlations to these could be causing the reduced variance of the largest eigenvalue.}
    \label{fig:largest_eigenvalue}
\end{figure}

The largest eigenvalue of the correlation matrix is a measure of the intensity of the correlation present, and in matrices inferred from financial returns tends to be significantly larger than the second largest \cite{laloux2000random} \cite{plerou2002random}. Generally this largest eigenvalue is larger during times of stress and smaller during times of calm \cite{drozdz2000440} \cite{laloux2000random}. Firstly we study how this varies over time for each correlation measure. This is shown in Figure \ref{fig:largest_eigenvalue}. For all of the networks there is a similar shape, with it peaking during times of market stress and dropping during times of calm. The Spearman and Pearson correlations have relatively similar values, although the Spearman has a smaller range. The $\tau$ correlation is much smaller than the other two at all times, and also has a smaller range. Times of stress and volatility tend to bring more outliers in returns data, which could be the cause of the difference in largest eigenvalue. Figure \ref{fig:correlation_comparison} has shown us that the $\tau$ correlation coefficients tend to be smaller than the others, which could explain why the largest eigenvalue is consistently smaller too. 

\subsection{MST Stability}

\begin{figure}
    \centering
        \begin{subfigure}{.30\textwidth}
          \includegraphics[width=\textwidth]{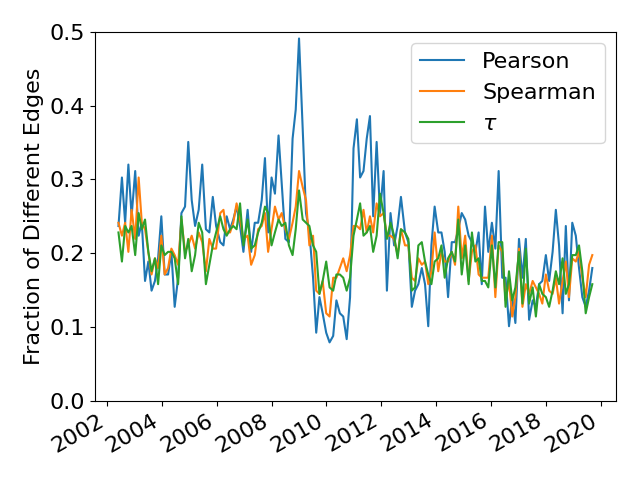}
          \caption{US}
        \label{fig:diff_us}
    \end{subfigure}
    \begin{subfigure}{.30\textwidth}
          \includegraphics[width=\textwidth]{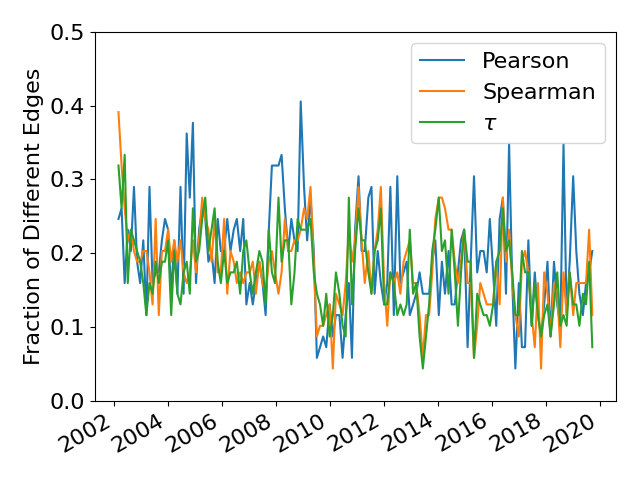}
          \caption{UK}
    \label{fig:diff_uk}
    \end{subfigure}
    \begin{subfigure}{.30\textwidth}
          \includegraphics[width=\textwidth]{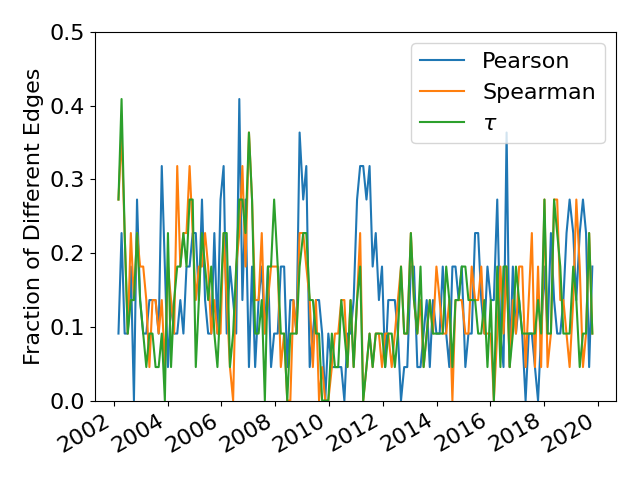}
          \caption{DE}
        \label{fig:diff_de}
    \end{subfigure}   
    \caption{Edge difference between adjacent MSTs. From this we can see the MSTs inferred using rank correlation are far more stable with regards to time than those inferred using Pearson correlation. While all of the trees seem to become more similar during the financial crisis (2009 - 2011), the Pearson MSTs show a big reconfiguration as the crisis starts, while the $\tau$ MSTs shows no spike before dropping.}
    \label{fig:edge_difference}
\end{figure}

From these inferred correlation matrices we transform them into distance matrices using (\ref{eq:dist}) and construct minimum spanning trees using Kruskal's algorithm. 
Firstly we focus on measuring the fraction of edge changes between MSTs adjacent in time, which quantifies how stable the trees are, and how well change in the market is detected. The results are plotted in Figure \ref{fig:edge_difference}. For the US and the UK, the Pearson MSTs tend to have much greater spikes in edge changes compared to the rank MSTs, indicating the rank MSTs are more robust to the outliers present during times of stress. Interestingly the number of edge changes seems to drop during the financial crisis in 2009 for all countries. Other authors have noted that by some measures the markets could be considered more stable during these times \cite{drozdz2000440} \cite{kocheturov2014523}, but we have not found that this has been mentioned in the context of MSTs. For Germany the results are harder to interpret, as the German MSTs are much smaller, meaning small changes are much more significant as an overall fraction and so we report the mean and standard deviation (s.d.) of these MST differences. The mean differences are are $0.138$ (s.d. $0.089$) for the Pearson MST, $0.131$ (s.d. $0.080$) for the Spearman and $0.128$ (s.d. $0.077$) for the $\tau$ MSTs. This indicates that the rank MSTs do change less than the Pearson ones, but the difference is small. 

This therefore shows that the Spearman and $\tau$ MSTs tend to be more stable than Pearson MSTs for all of the countries. This is particularly prominent at the start of the financial crisis for the US and the UK, with the Pearson MSTs showing a large spike in difference, while the Spearman and $\tau$ MSTs show little or no change in difference. In this particular situation we would expect the heavy tails to affect the Pearson correlation between two assets more than the rank methods, and this should change the edges selected by the MST construction procedure. 

\begin{figure}
    \centering
     \begin{subfigure}{.30\textwidth}
          \includegraphics[width=\textwidth]{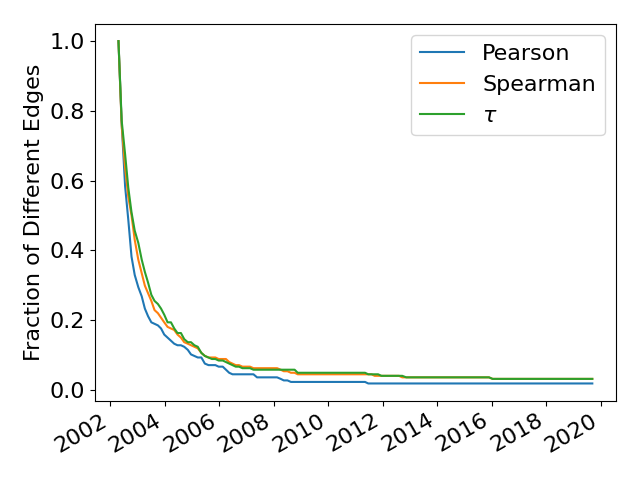}
          \caption{US}
        \label{fig:edges_life_us}
    \end{subfigure}
    \begin{subfigure}{.30\textwidth}
          \includegraphics[width=\textwidth]{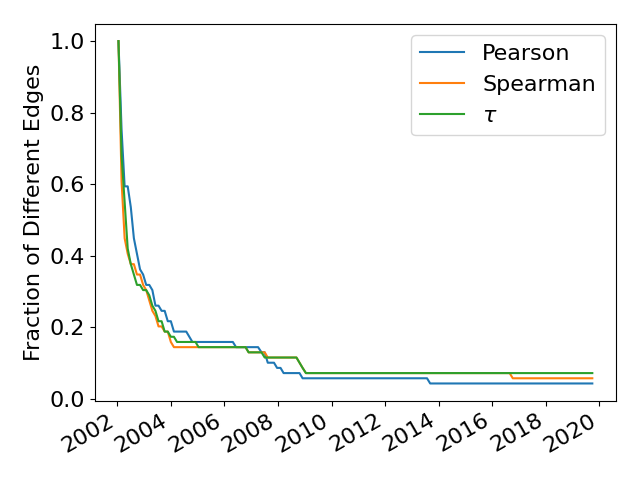}
          \caption{UK}
    \label{fig:edges_life_uk}
    \end{subfigure}
    \begin{subfigure}{.30\textwidth}
          \includegraphics[width=\textwidth]{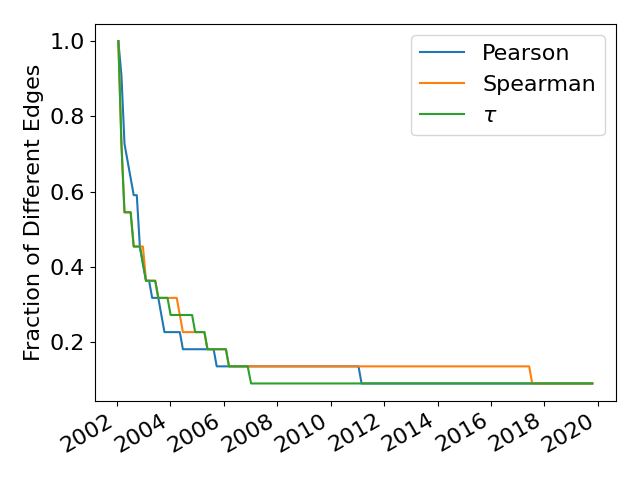}
          \caption{DE}
        \label{fig:edges_life_de}
    \end{subfigure}   
    \caption{Multi-step survival ratio for the MSTs. This gives us a measure of how long the edges persist for. Most of the edges disappear very rapidly, with around 70\% changing within 2 years. The rank MSTs seem to be slightly more stable than the Pearson MSTs, maintaining more edges from the initial tree.}
    \label{fig:tree_diff_from_first}
\end{figure}

Next we measure how the MSTs have changed from the first inferred tree using the multi step survival ratio \cite{onnela2003dynamics} - the fraction of edges that have been consistently maintained for each tree from the first to the current. This measures the life of an edge and shows us how the tree evolves. This is defined as
\begin{equation}
    \frac{1}{p-1} |E(t) \cap E(t-1) \dots E(t-k+1) \cap E(t-k)|    
\end{equation}
where $E(t)$ is the edge set at that moment in time, $k$ goes from 1 to $t-1$ and $|S|$ is the cardinality of the set $S$.
A plot of this is shown in Figure \ref{fig:tree_diff_from_first}. From this we can see that quite rapidly the trees differ from the original for all countries, with 70\% of the edges changing within 2 years. For our experiments, what is particularly interesting is that for the US and the UK the rank MSTs maintain slightly more edges than the Pearson MSTs, but the difference between the $\tau$ and Spearman MSTs is very small. For Germany, the Spearman MSTs actually maintain the most edges for the longest time period, followed by the Pearson MSTs and then the $\tau$ MSTs.

Over the entire dataset for the US, the Pearson MSTs maintain 4 edges, the Spearman MSTs 7 edges and the $\tau$ MSTs 8 edges. For the UK the Pearson MSTs maintain 3 edges, the Spearman MSTs 5 edges and the $\tau$ MSTs 4 edges. For Germany all three maintain 2 edges. All of these edges are intrasector aside from one in the German $\tau$ MSTs, which is BASF to Bayer (Materials to Healthcare).

\begin{figure}
    \centering
    \begin{subfigure}{.30\textwidth}
          \includegraphics[width=\textwidth]{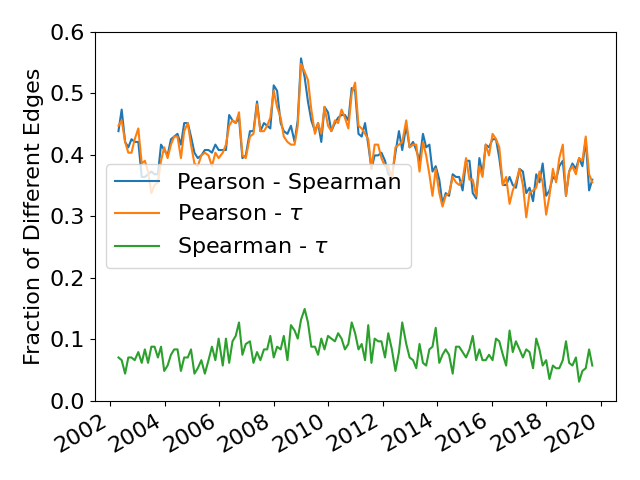}
          \caption{US}
        \label{fig:edge_diff_us}
    \end{subfigure}
    \begin{subfigure}{.30\textwidth}
          \includegraphics[width=\textwidth]{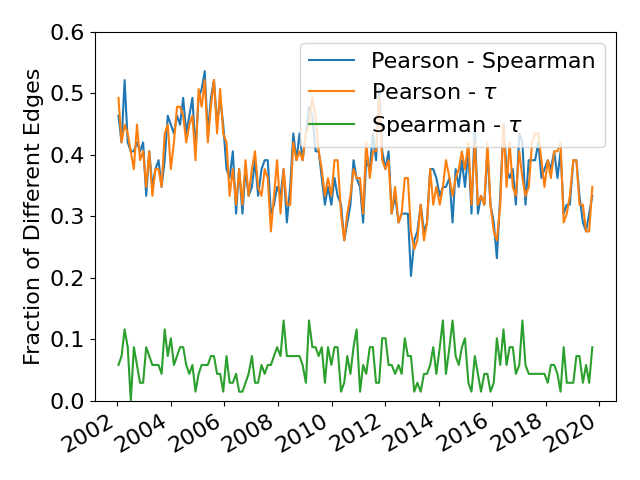}
          \caption{UK}
    \label{fig:edge_diff_uk}
    \end{subfigure}
    \begin{subfigure}{.30\textwidth}
          \includegraphics[width=\textwidth]{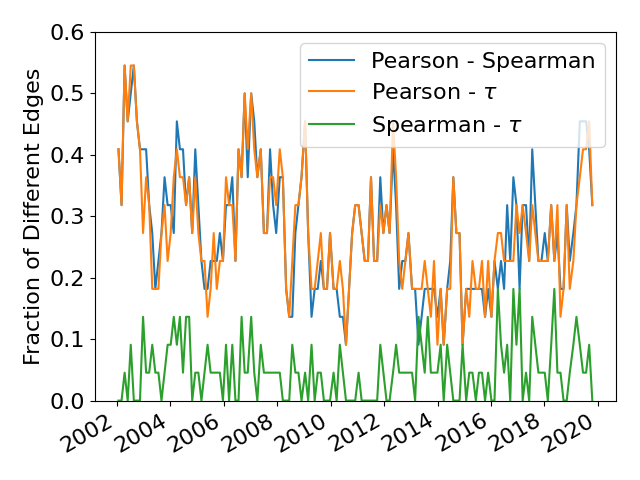}
          \caption{DE}
        \label{fig:edge_diff_de}
    \end{subfigure}   
    \caption{Fraction of the edges that differ between the various MSTs. The US has the largest difference, followed by the UK and then Germany, which indicates size of the MST influences the edge difference. In general for the US and the UK around 40\% of the edges differ, although this does increase during the financial crisis, particularly in 2009. Since the German tree is so much smaller there is a larger range in these values, but on average around 30\% of the edges different between the Pearson and rank MSTs. }
    \label{fig:pearson_spearman_difference}
\end{figure}

There is of course the question of how the difference between the MSTs changes over time. We measure the fraction of edges that differ between the three MSTs and plot it in Figure \ref{fig:pearson_spearman_difference}. From this we can see there is a significant difference in the presence of edges between the rank MSTs and the Pearson MSTs for all countries. The difference does seem to increase during the banking and financial crises, with notable peaks occurring during 2008 and 2009. There seems to be relatively little difference between the two rank MSTs, with less than 10\% of the edges being different for every country for most of the time period. This difference between the rank methods does not seem to be particularly affected by market conditions.

\subsection{Node and Sector Centrality}

    These networks can be used to study the importance of economic sectors in the markets of the selected countries, and how they might be implicated in various crises and growth periods that occur. Here we are interested in how sectors may be differently expressed in the MSTs, to see if the different correlation coefficients can give an alternative picture of the market at that moment. 

\begin{figure}[H]
    \centering%
    \begin{minipage}{0.95\textwidth}
        \begin{subfigure}{0.46\textwidth}
         \includegraphics[width=\textwidth]{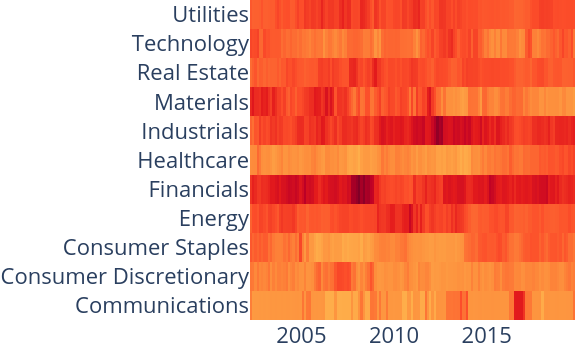}
        \caption{US - Pearson}
    \end{subfigure}
    \begin{subfigure}{0.25\textwidth}
         \includegraphics[width=\textwidth]{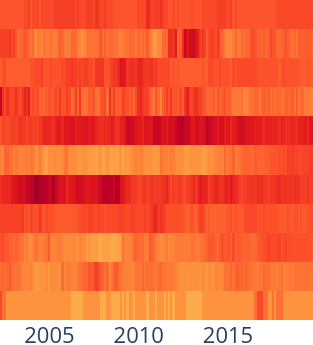}
        \caption{US - Spearman}
    \end{subfigure}
        \begin{subfigure}{0.25\textwidth}
         \includegraphics[width=\textwidth]{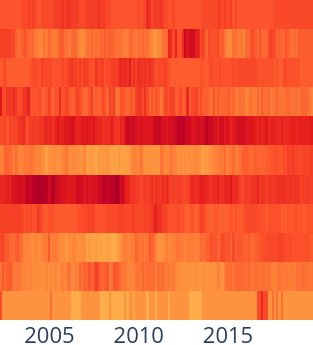}
        \caption{US - $\tau$}
        \end{subfigure}
    \end{minipage}%
    \begin{minipage}{0.05\textwidth}
        \includegraphics[width=\textwidth, trim=0cm 0 0 2.5cm]{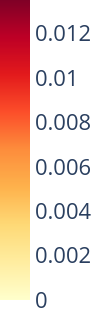}
    \end{minipage}
    \label{fig:sector_degree_centrality_US}
    \begin{minipage}{0.95\textwidth}
        \begin{subfigure}{0.44\textwidth}
         \includegraphics[width=\textwidth]{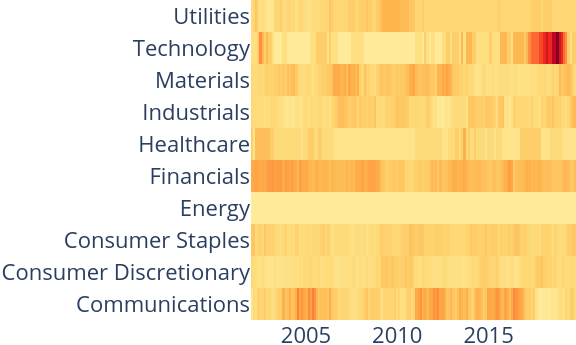}
        \caption{UK - Pearson}
    \end{subfigure}
    \begin{subfigure}{0.25\textwidth}
         \includegraphics[width=\textwidth]{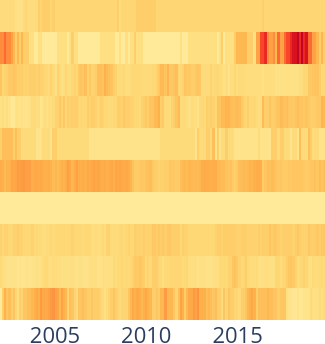}
        \caption{UK - Spearman}
    \end{subfigure}
        \begin{subfigure}{0.25\textwidth}
         \includegraphics[width=\textwidth]{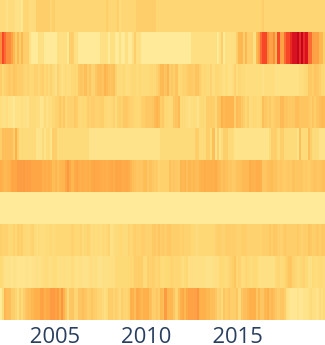}
        \caption{UK - $\tau$}
        \end{subfigure}
    \end{minipage}%
    \begin{minipage}{0.05\textwidth}
        \includegraphics[width=\textwidth, trim=-0cm 0 0 2.5cm]{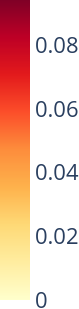}
    \end{minipage}
    \begin{minipage}{0.95\textwidth}
        \begin{subfigure}{0.44\textwidth}
         \includegraphics[width=\textwidth]{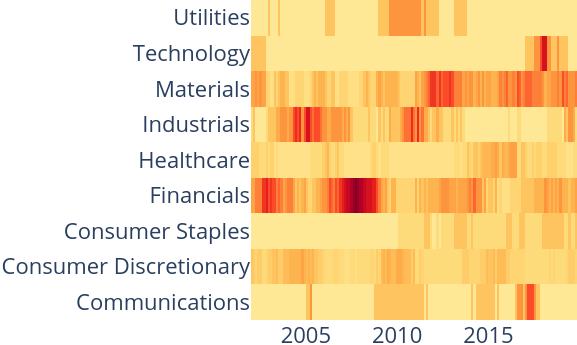}
        \caption{DE - Pearson}
    \end{subfigure}
    \begin{subfigure}{0.25\textwidth}
         \includegraphics[width=\textwidth]{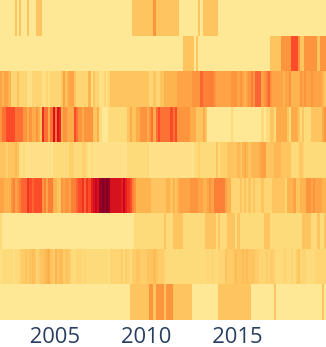}
        \caption{DE - Spearman}
    \end{subfigure}
        \begin{subfigure}{0.25\textwidth}
         \includegraphics[width=\textwidth]{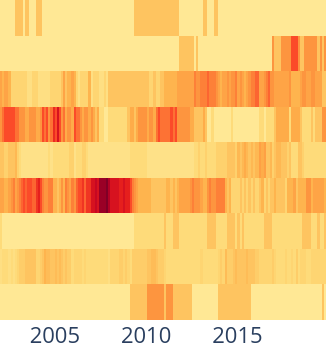}
        \caption{DE - $\tau$}
        \end{subfigure}
    \end{minipage}%
    \begin{minipage}{0.05\textwidth}
        \includegraphics[width=\textwidth, trim=-0cm 0 0 2.5cm]{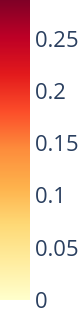}
    \end{minipage}
    \caption{Degree centrality for each of the sectors over time}
    \label{fig:sector_degree_centrality}
\end{figure}

\begin{figure}[H]
    \centering%
    \begin{minipage}{0.95\textwidth}
        \begin{subfigure}{0.45\textwidth}
         \includegraphics[width=\textwidth]{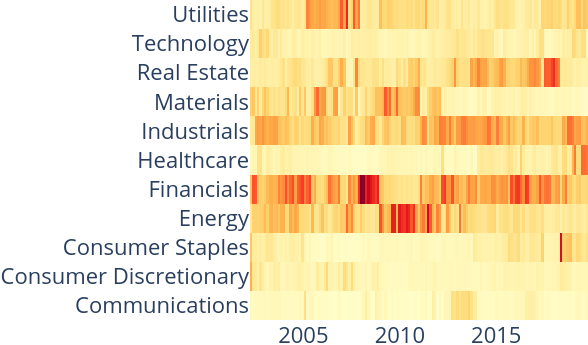}
        \caption{US - Pearson}
    \end{subfigure}
    \begin{subfigure}{0.25\textwidth}
         \includegraphics[width=\textwidth]{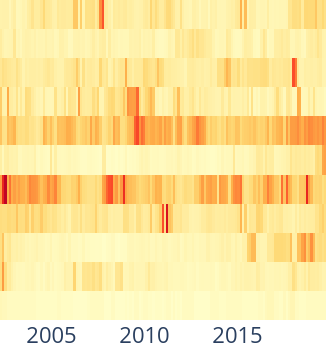}
        \caption{US - Spearman}
    \end{subfigure}
        \begin{subfigure}{0.26\textwidth}
         \includegraphics[width=\textwidth]{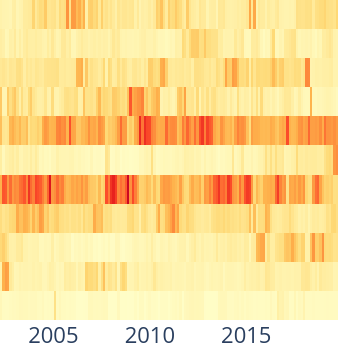}
        \caption{US - $\tau$}
        \end{subfigure}
    \end{minipage}%
    \begin{minipage}{0.05\textwidth}
        \includegraphics[width=\textwidth, trim=-0cm 0 0 2.5cm]{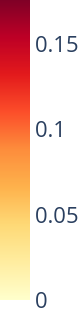}
    \end{minipage}
    \begin{minipage}{0.95\textwidth}
        \begin{subfigure}{0.44\textwidth}
         \includegraphics[width=\textwidth]{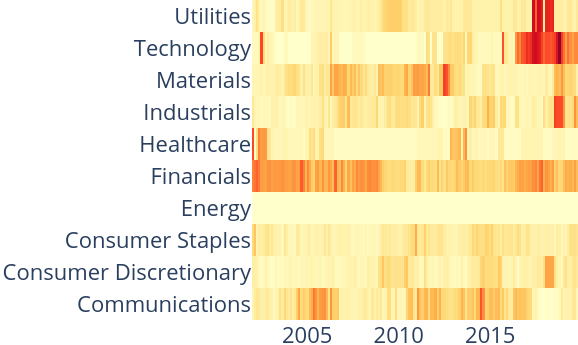}
        \caption{UK - Pearson}
    \end{subfigure}
    \begin{subfigure}{0.25\textwidth}
         \includegraphics[width=\textwidth]{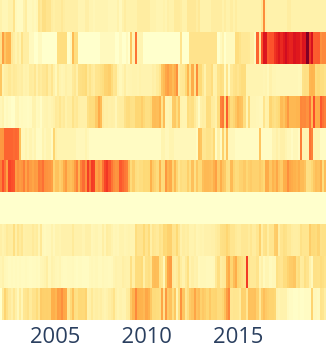}
        \caption{UK - Spearman}
    \end{subfigure}
        \begin{subfigure}{0.25\textwidth}
         \includegraphics[width=\textwidth]{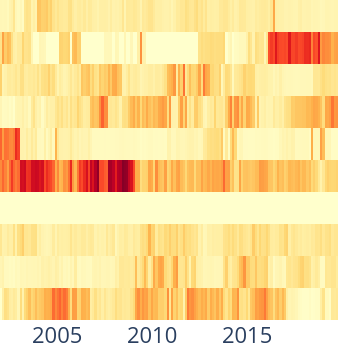}
        \caption{UK - $\tau$}
        \end{subfigure}
    \end{minipage}%
    \begin{minipage}{0.05\textwidth}
        \includegraphics[width=\textwidth, trim=-0cm 0 0 2.5cm]{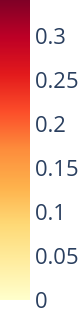}
    \end{minipage}
    \centering%
    \begin{minipage}{0.95\textwidth}
        \begin{subfigure}{0.44\textwidth}
         \includegraphics[width=\textwidth]{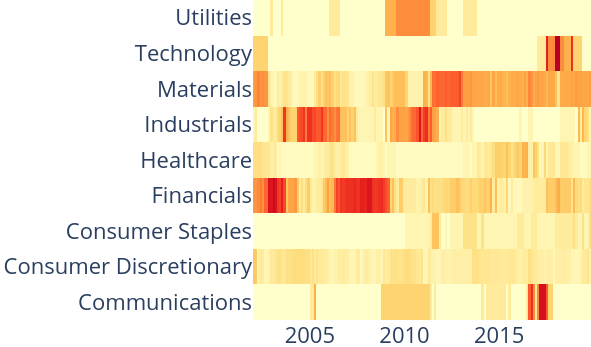}
        \caption{DE - Pearson}
    \end{subfigure}
    \begin{subfigure}{0.25\textwidth}
         \includegraphics[width=\textwidth]{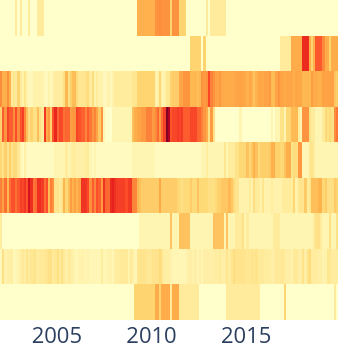}
        \caption{DE - Spearman}
    \end{subfigure}
        \begin{subfigure}{0.25\textwidth}
         \includegraphics[width=\textwidth]{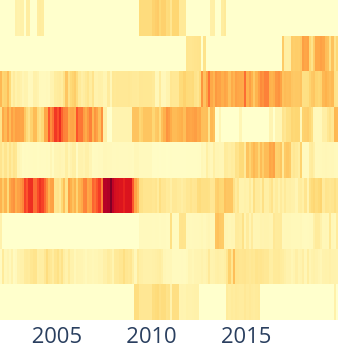}
        \caption{DE - $\tau$}
        \end{subfigure}
    \end{minipage}%
    \begin{minipage}{0.05\textwidth}
        \includegraphics[width=\textwidth, trim=-0cm 0 0 2.5cm]{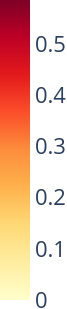}
    \end{minipage}
    \caption{Betweenness centrality for each sector over time}
    \label{fig:sector_betweenness_centrality}
\end{figure}

 To measure this we look at how the centrality of a sector varies by calculating the mean of the centrality of all the nodes in said sector. This reduces the effect of the different numbers of companies in each sector. Then to make comparisons between the differently sized MSTs easier, we normalize the sector centrality so the sum of all sector centralities in an MST is 1. To express mathematically, we calculate the centrality of sector $s$ from the set of all sectors $S$ as follows
    \begin{equation}
        \frac{1}{\sum_{j \in S} C_j}  \frac{1}{|s|} \sum_{i \in s} c_i
    \end{equation}
    where $c_i$ is the centrality of node $i$, and $|s|$ is the cardinality of set $s$. 
    We use both unweighted degree centrality and betweenness centrality to measure the influence of a sector. Betweenness centrality is calculated by looking at the fraction of shortest paths that pass through a node, and allows us to get a different perspective on which nodes are more important. The results are shown in Figures \ref{fig:sector_degree_centrality} (degree centrality) and \ref{fig:sector_betweenness_centrality} (betweenness centrality). 
    
    Firstly we focus on the sector degree centrality. For the US, the Financials and Industrials sectors are important in all three MSTs for the entire dataset. They also seem relatively similarly expressed over time in all three. The Materials sector is regarded as more important at the start of the Pearson MSTs (from 2003 - 2008) but less so in the rank MSTs. 
    The Technology sector becomes more important in the rank MSTs than in the Pearson MSTs, while the Energy sector is slightly more important in the Pearson MSTs from 2009 - 2010 compared to the rank MSTs. The Consumer Discretionary, Healthcare and Utilities sectors all are relatively similar throughout the dataset.
    
    For the UK again the Financials sector is important in all three MSTs. The most notable feature is the large spike in importance of the Technology sector from 2015, which is present in all three MSTs again. The Utilities sector seems more important in the Pearson MSTs than the rank ones, while the Industrials sector seems slightly more important in the rank MSTs, notably from 2015 onwards. 
    
    For Germany the Financials, Industrials and Materials sectors are again important in all three MSTs. The Industrials sector seems more important in the rank MSTs, notably from the start of the dataset until 2008 and then from 2016 onwards. The Communications sector is quite differently expressed between the Pearson and rank MSTs, but is very small, so the effects of one company are large. The spike in the centrality of the Technology sector in 2017 is more intense in the Pearson MSTs than the rank ones, but the rest of the centrality seems similar. 
    
    In general it seems the degree centrality of most sectors is relatively similar in all three MSTs, with only small differences occurring. Furthermore the sector centralities of the rank MSTs are virtually identical. 


Next we study the betweenness centralities of the sectors. Starting with the US, again the Industrials and Financials sectors are central for the majority of the dataset in all three MSTs. Both seem more central in the $\tau$ MSTs than the Pearson or Spearman. The Energy sector becomes notably central in the Pearson MSTs from 2009 - 2012 but this is not reflected in the rank MSTs, and there is also a rise in centrality for the Utilities sector in the Pearson MSTs that again is not seen in the rank MSTs. The Consumer Staples sector seems more central in the rank MSTs, notably towards the end of the dataset from 2015 onwards, as does the Consumer Discretionary sector. 

For the UK the Financials sector is again central throughout the dataset in all three MSTs, but has a higher centrality in the $\tau$ MSTs. The Technology sector also becomes very central in all three from 2016, but has the largest centrality in the Pearson MSTs. The Utilities sector is not central for most of the dataset in any of the MSTs, but has a peak during 2016 for the Pearson MSTs, but not in the rank MSTs. The Communications and Consumer Staples sectors are relatively similarly expressed across all MSTs, and are not very central. The centrality of the Materials sector varies similarly in all MSTs, but has a higher peak value in the Pearson MSTs, and the Consumer Discretionary is similar, but has its peak value in the Spearman MSTs. 

For Germany the Financials, Industrials and Materials sectors are regarded as important throughout most of the dataset for every MST. However the Materials and Industrials sectors are more central in the Pearson and Spearman MSTs than the $\tau$. At the end of the time period, the Technology sector has a higher peak in the Pearson MSTs, followed by the Spearman MSTs, but is not particularly important in the $\tau$ MSTs. The Consumer Staples, Consumer Discretionary and Healthcare sectors are not particularly central in any of the MSTs, and are relatively similarly expressed. 

Compared to the sector degree centrality, the sector betweenness centrality shows a much greater range, and greater disagreement between the MSTs, notably between the rank MSTs. However in general the sectors that are regarded as important in the degree MSTs are also regarded as important in the betweenness MSTs. This could imply that companies tend to be placed in different positions in the different MSTs, even if they have a similar degree centrality 

\subsection{Effects of Gaussian Deviations}
\label{sec:mst_exploration}
\begin{figure}
\centering
    \begin{subfigure}{.3\textwidth}
  \includegraphics[width=\textwidth]{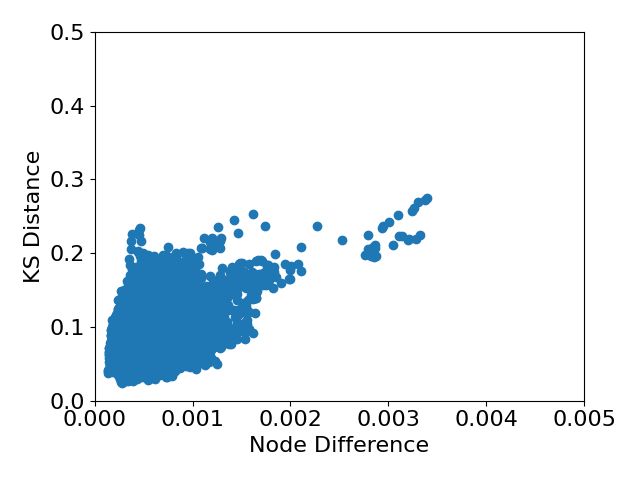}
  \caption{US Pearson - Spearman}
    \end{subfigure}%
        \begin{subfigure}{.3\textwidth}
  \includegraphics[width=\textwidth]{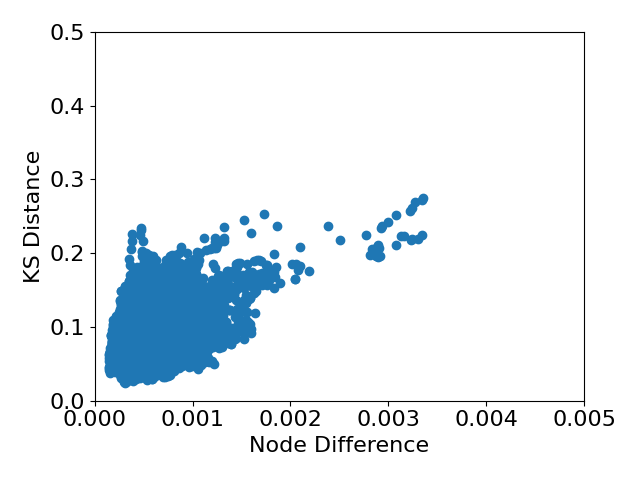}
  \caption{US Pearson - $\tau$}
    \end{subfigure}%
        \begin{subfigure}{.3\textwidth}
          \includegraphics[width=\textwidth]{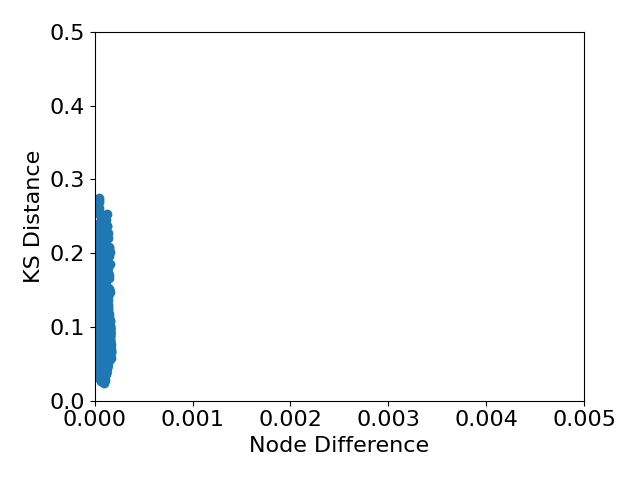}
  \caption{US Spearman - $\tau$}
    \end{subfigure}
    
    \begin{subfigure}{.3\textwidth}
      \includegraphics[width=\textwidth]{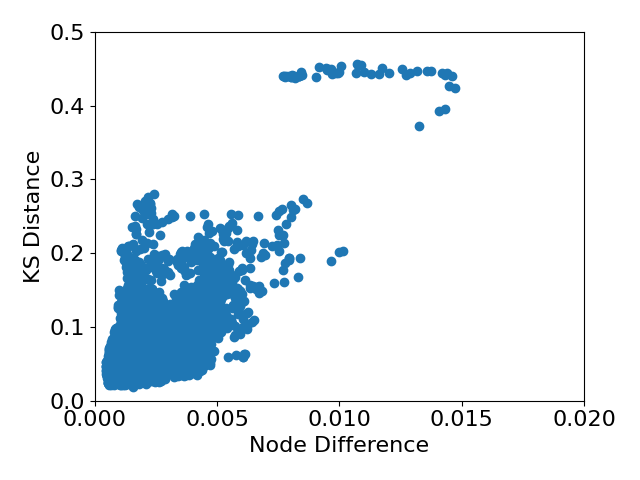}
  \caption{UK Pearson - Spearman}
    \end{subfigure}%
        \begin{subfigure}{.3\textwidth}
  \includegraphics[width=\textwidth]{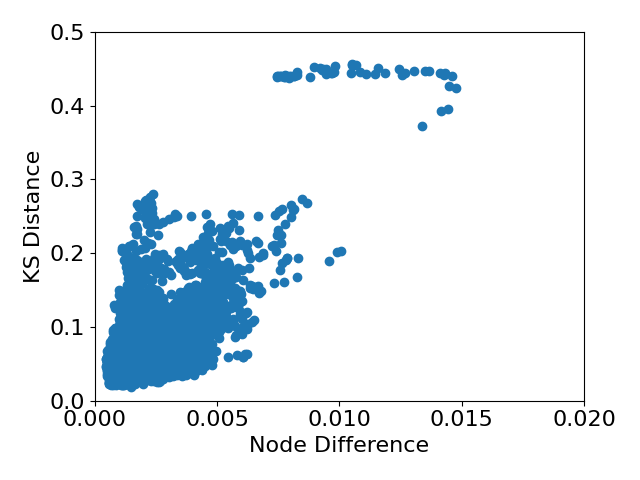}
  \caption{UK Pearson - $\tau$}
    \end{subfigure}%
        \begin{subfigure}{.3\textwidth}
          \includegraphics[width=\textwidth]{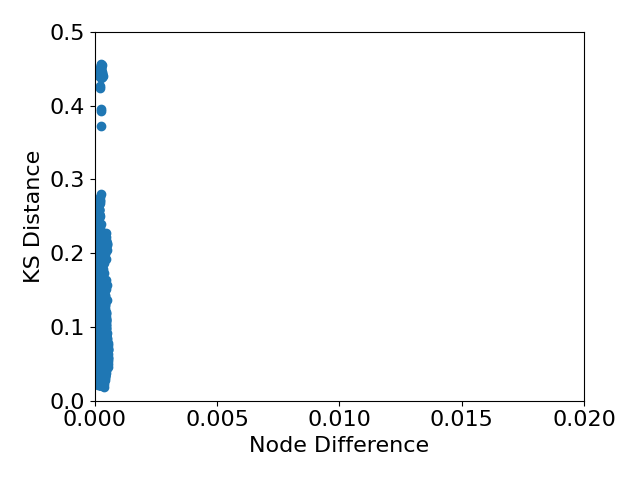}
  \caption{UK Spearman - $\tau$}

    \end{subfigure}
    
    \begin{subfigure}{.3\textwidth}
  \includegraphics[width=\textwidth]{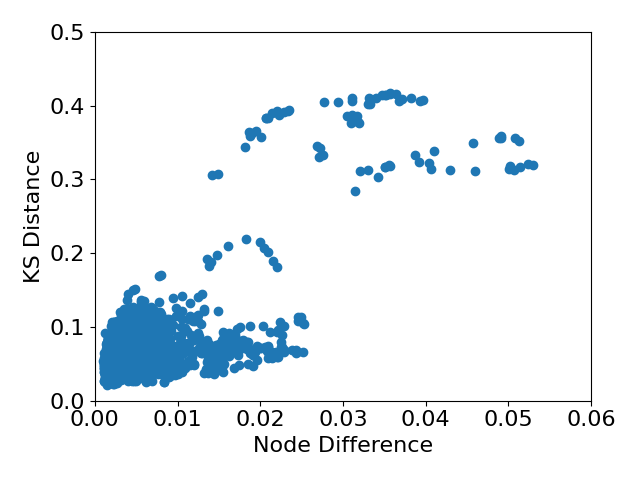}
  \caption{Germany Pearson - Spearman}
    \end{subfigure}%
        \begin{subfigure}{.3\textwidth}
  \includegraphics[width=\textwidth]{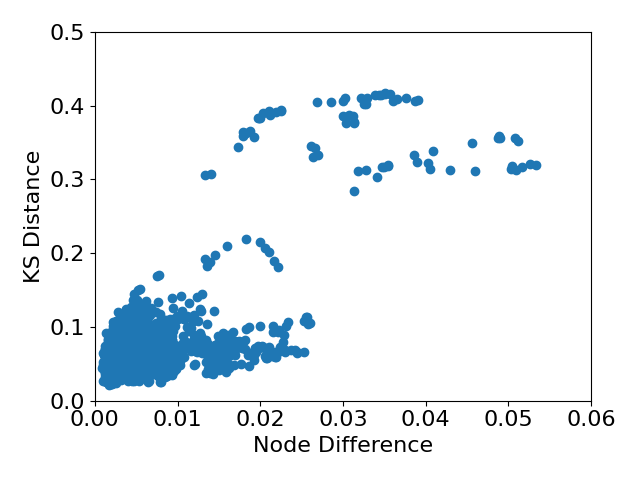}
  \caption{Germany Pearson - $\tau$}
    \end{subfigure}%
        \begin{subfigure}{.3\textwidth}
  \includegraphics[width=\textwidth]{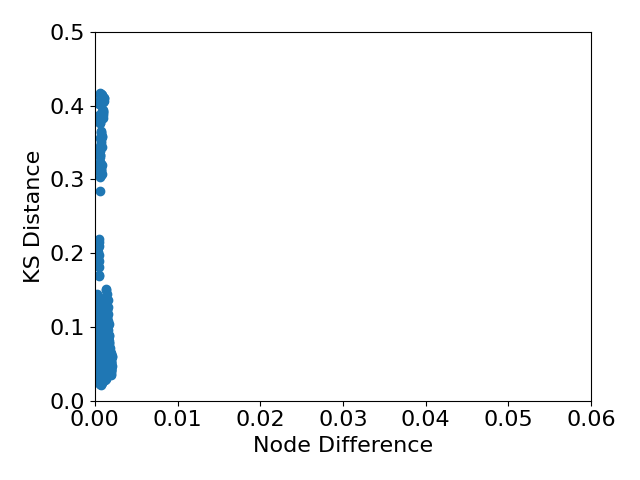}
  \caption{Germany Spearman - $\tau$}
    \end{subfigure}%
    \caption{Scatter plots of the difference for nodes in the full correlation matrix against the KS distance for all 3 countries. There appears to be a positive relationship between node difference and KS distance when comparing the Pearson and rank MSTs, indicating that a deviation from univariate Gaussianity does cause differences }
    \label{fig:full_correlation_node_difference}
\end{figure}

\begin{figure}
\centering
    \begin{subfigure}{.3\textwidth}
  \includegraphics[width=\textwidth]{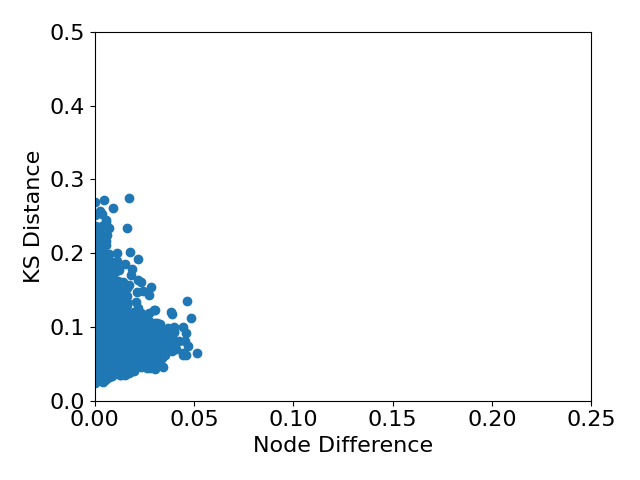}
  \caption{US Pearson - Spearman}
    \end{subfigure}%
        \begin{subfigure}{.3\textwidth}
  \includegraphics[width=\textwidth]{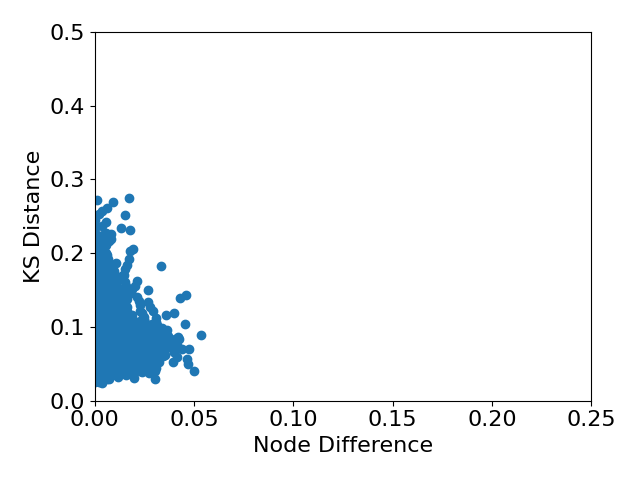}
  \caption{US Pearson - $\tau$}
    \end{subfigure}%
        \begin{subfigure}{.3\textwidth}
          \includegraphics[width=\textwidth]{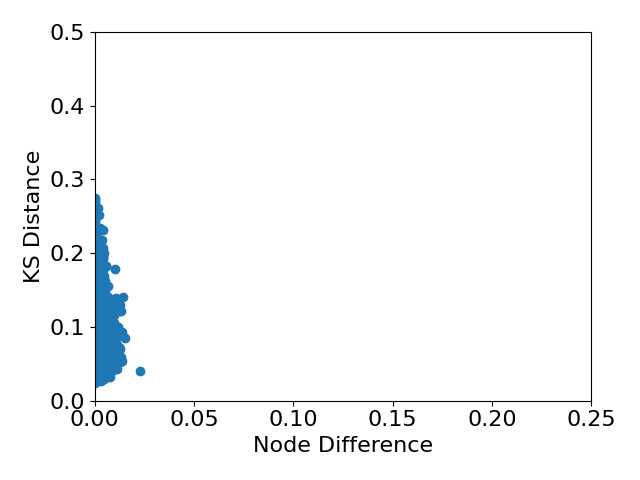}
  \caption{US Spearman - $\tau$}
    \end{subfigure}
    \begin{subfigure}{.3\textwidth}

    \includegraphics[width=\textwidth]{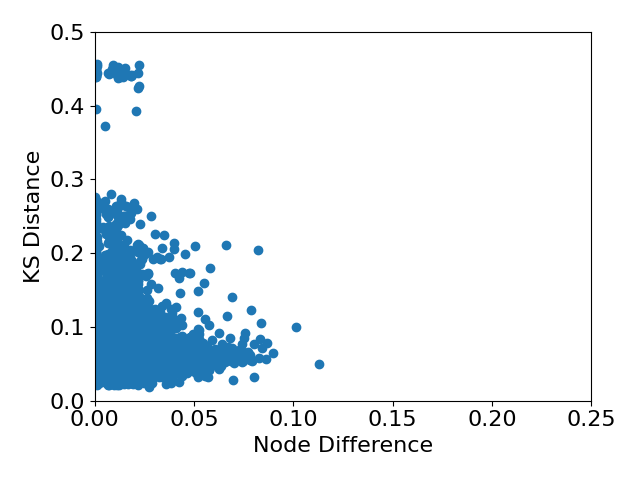}
  \caption{UK Pearson - Spearman}
    \end{subfigure}%
        \begin{subfigure}{.3\textwidth}
  \includegraphics[width=\textwidth]{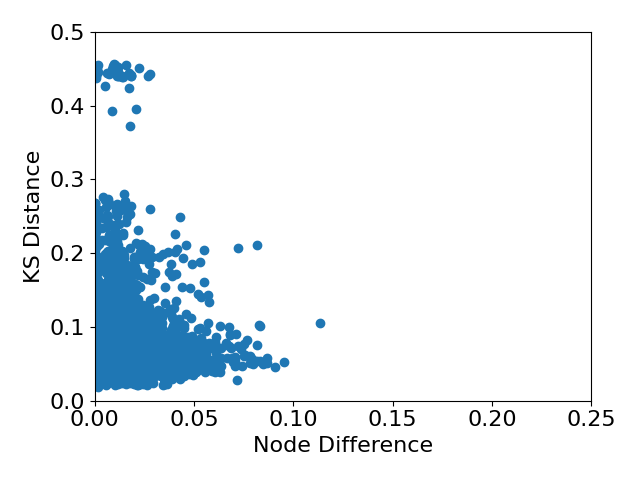}
  \caption{UK Pearson - $\tau$}
    \end{subfigure}%
        \begin{subfigure}{.3\textwidth}
  \includegraphics[width=\textwidth]{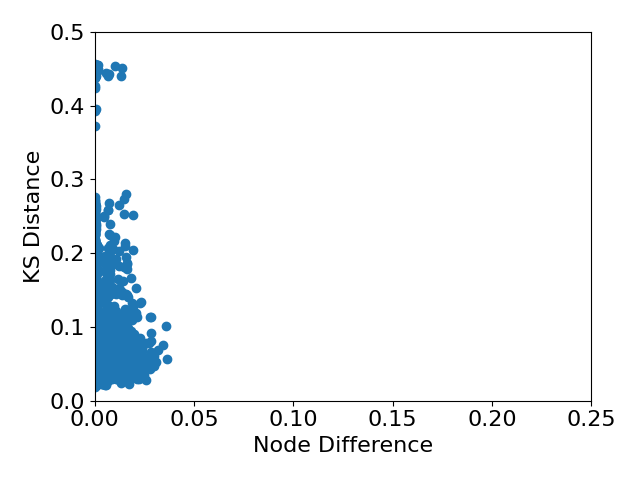}
  \caption{UK Spearman - $\tau$}
    \end{subfigure}
    \begin{subfigure}{.3\textwidth}
          \includegraphics[width=\textwidth]{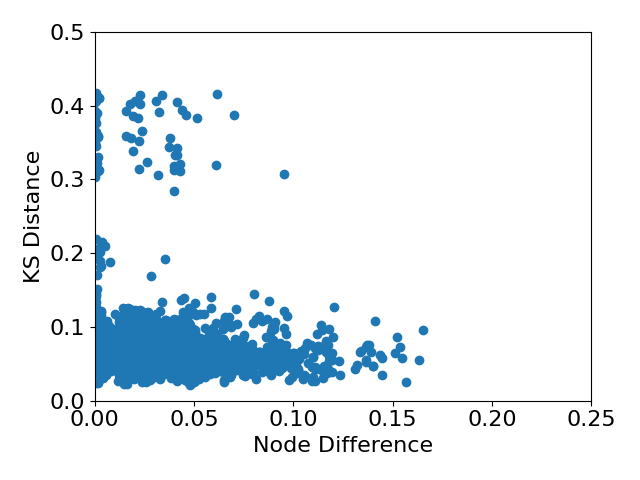}
  \caption{Germany Pearson - Spearman}
    \end{subfigure}
        \begin{subfigure}{.3\textwidth}
    \includegraphics[width=\textwidth]{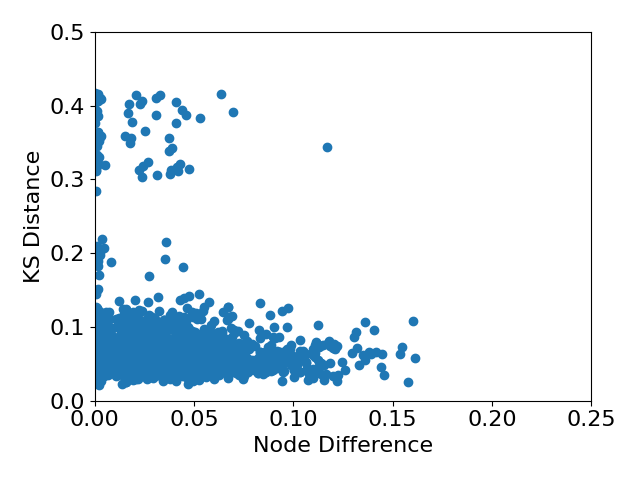}
  \caption{Germany Pearson - $\tau$}
    \end{subfigure}%
        \begin{subfigure}{.3\textwidth}
  \includegraphics[width=\textwidth]{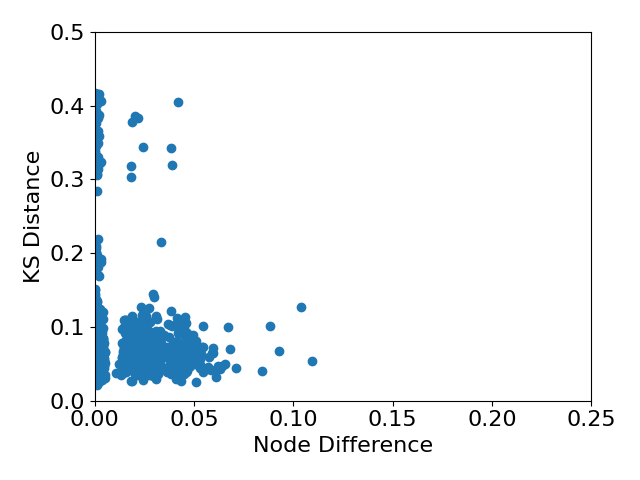}
  \caption{Germany Spearman - $\tau$}
    \end{subfigure}%
    \caption{Scatter plots of the difference for nodes in the MSTs against the KS distance for all 3 countries. There seems to be significantly less of a relationship between the node difference and the KS distance in the MSTs compared to the full correlation matrices. This could be due to the MST procedure discarding the changed relationships.}
    \label{fig:mst_node_difference}
\end{figure}

In this section we explore the potential reasons for differences between the MSTs, focusing on univariate non-Gaussianity. To measure this we use the Kolmogorov-Smirnov (KS) distance from a Gaussian distribution for each stock for each window and plot this against the absolute normalised node difference, measured by
\begin{equation}
\sum_{i=1}^p \sum_{j=i+1}^p |\frac{C_i^x}{M^x} - \frac{C_i^y}{M^y}|    
\end{equation}
where $M^x = \sum_{i=1}^p \sum_{j=1}^p C_{ij}$ (i.e. the sum of all the correlations in the network), $C_i$ is the $i$th column of the correlation matrix and $C^x$ and $C^y$ are correlation matrices created from different correlation coefficients. Normalising the correlations ensures that times where the correlations are higher do not distort our measurements. This is done for both the full correlation matrices (Figure \ref{fig:full_correlation_node_difference}) and the MST filtered correlation matrices (Figure \ref{fig:mst_node_difference}). To clarify, an MST filtered correlation matrix is the correlation matrix constructed from the MST, where edges in the MST are given the weight of their correlation from the original full correlation matrix and all other correlations are set to zero. 

For all countries there is a positive relationship between the KS distance and the distance between nodes in the full correlation matrices when we compare the Pearson and rank networks, indicating that a departure from Gaussianity does increase the difference between the different correlation coefficients. There appears to be no relationship when comparing the rank correlation networks, which is to be expected. 

However if we look at the MST figures, the results seem a bit different. For all countries it seems there is relatively little relationship between node difference and KS distance. However since the procedure to create the MST discards the majority of correlations, this could imply that the trees tend to keep correlations that are unaffected by this deviation. Furthermore, since the MSTs only keep large correlations, it could be that the deviations affect smaller correlations more. 

To quantify these we show the Spearman correlation between node difference and KS distance in Table \ref{tab:node_vs_ks_distance}. For all of the full correlation matrices there is some positive correlation between deviation from Gaussianity and the difference between the rank and Pearson correlations. The US and UK also have some very mild negative correlation between the $\tau$ and Spearman MSTs. None of the countries have any correlation between the node difference and KS distance with the MSTs. From the scatter plots this is mostly to be expected. The results if we use unweighted MSTs are very similar.

\begin{table}[]
    \centering
    \begin{tabular}{|c|c|c|c|}
        \hline
        Network & Pearson - Spearman & Pearson - $\tau$ & Spearman - $\tau$ \\
        \hline
        US Full & 0.221 & 0.244 & -0.114 \\
        UK Full & 0.409 & 0.415 & -0.104 \\
        DE Full & 0.331 & 0.352 & -0.075 \\
        \hline
        US MST & -0.005 & -0.005 & -0.016 \\
        UK MST & 0.000 & -0.008 & -0.006 \\
        DE MST & -0.013 & 0.008 & 0.021 \\
        \hline
    \end{tabular}
    \caption{Spearman correlation between node difference and the Kolmogorov-Smirnov distance for the node from a univariate Gaussian. In general a departure from univariate Gaussianity tends to cause differences in the full correlation matrices, but not in the MSTs, potentially due to the filtration procedure.}
    \label{tab:node_vs_ks_distance}
\end{table}

\begin{figure}
    \centering
    \begin{subfigure}{0.3\textwidth}
      \includegraphics[width=\textwidth]{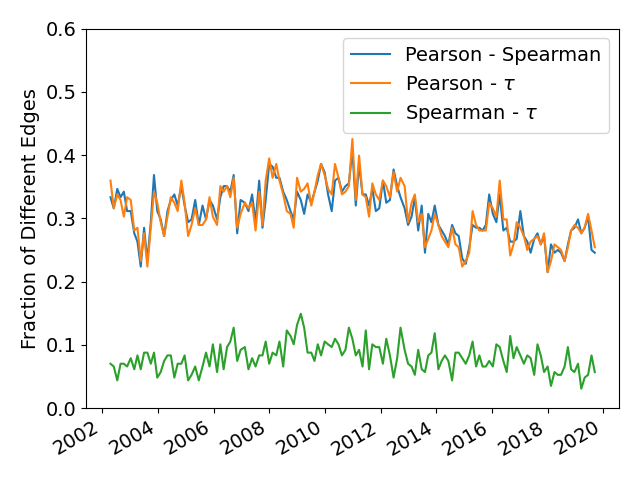}
      \caption{US}
    \end{subfigure}
    \begin{subfigure}{0.3\textwidth}
      \includegraphics[width=\textwidth]{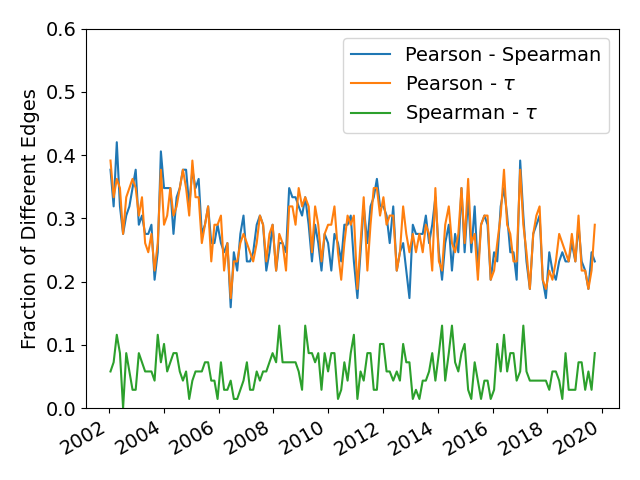}
      \caption{UK}
    \end{subfigure}
    \begin{subfigure}{0.3\textwidth}
      \includegraphics[width=\textwidth]{edge_difference_quantile_UK.png}
      \caption{DE}
    \end{subfigure}
    \caption{Edge difference between the trees over time when quantile normalisation is used to make the asset returns data normal. Comparing this to Figure \ref{fig:edge_difference} there is a reduction in this difference between the Pearson and rank MSTs, but it is still much higher than the difference between the rank MSTs. This would imply that it is not just a deviation from normality which causes differences between the MSTs.}
    \label{fig:edge_difference_quantile}
\end{figure}

Our second experiment on this front is to use quantile normalisation to make the distributions of the asset returns normal. We then look at how this changes the differences between the trees. We use 200 quantiles for this, and plot the differences between the MSTs in Figure \ref{fig:edge_difference_quantile}. If we compare this to Figure \ref{fig:edge_difference} we can see the differences between the Pearson and rank methods have been reduced, but that they are still larger than the differences between the rank MSTs. 

This therefore implies that, in contrast to the previous results, the departure from univariate Gaussianity does cause differences between the MSTs as well as the full correlation matrices. However it does not explain all of the differences between the MSTs. This would imply that overall there are non-linear relationships present in the dataset that also drive differences between the MSTs, as well as non-normalities.  

\subsection{MST Topology}

\begin{figure}
\centering
    \begin{subfigure}{.3\textwidth}
  \includegraphics[width=\textwidth]{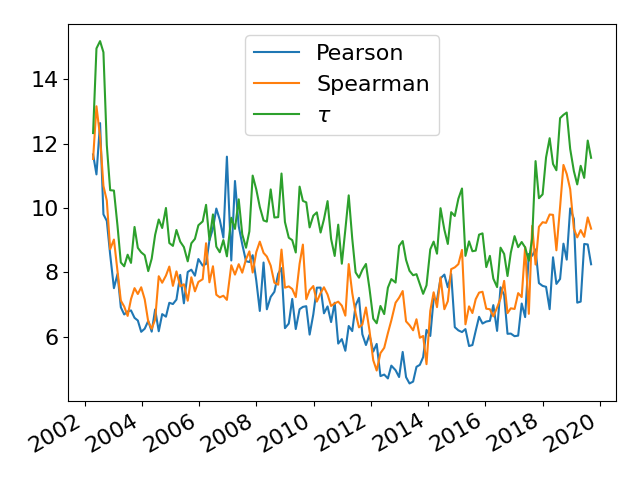}
  \caption{US}
    \end{subfigure}
        \begin{subfigure}{.3\textwidth}
  \includegraphics[width=\textwidth]{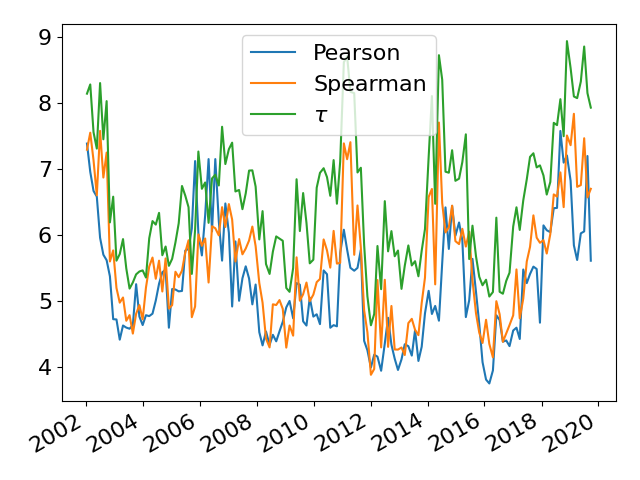}
  \caption{UK}
    \end{subfigure}%
        \begin{subfigure}{.3\textwidth}
  \includegraphics[width=\textwidth]{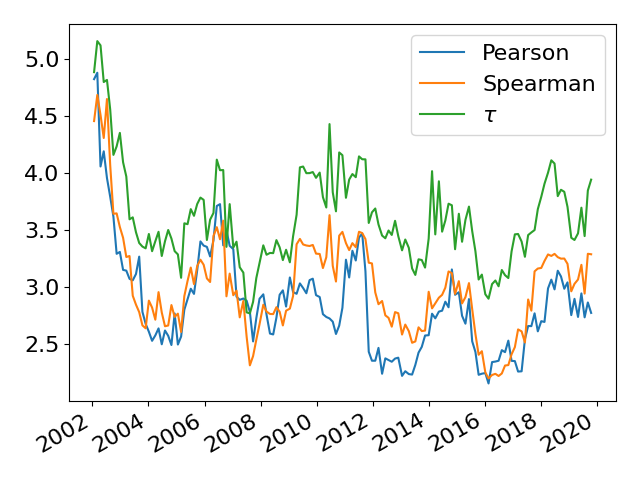}
  \caption{DE}
    \end{subfigure}%
    \caption{Average Shortest Path Length}
    \label{fig:av_shortest_path}

    \begin{subfigure}{.3\textwidth}
  \includegraphics[width=\textwidth]{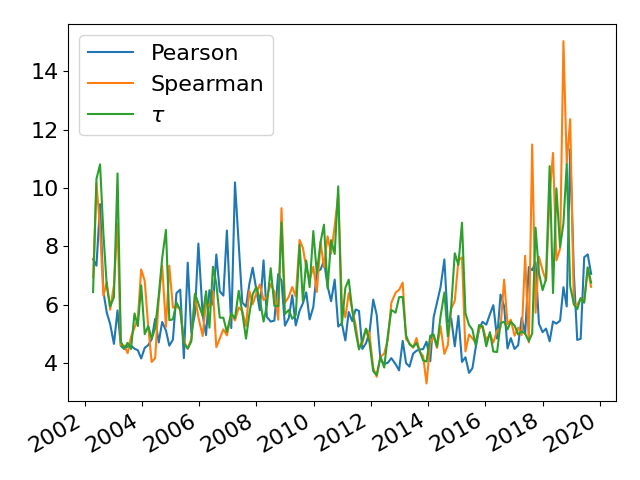}
  \caption{US}
    \end{subfigure}
        \begin{subfigure}{.3\textwidth}
  \includegraphics[width=\textwidth]{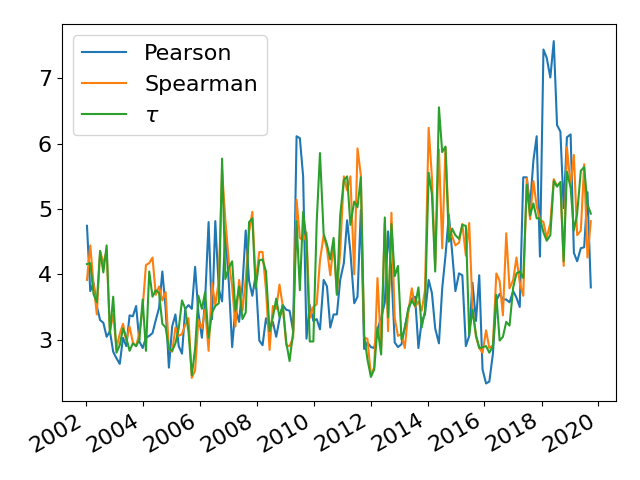}
  \caption{UK}
    \end{subfigure}%
        \begin{subfigure}{.3\textwidth}
  \includegraphics[width=\textwidth]{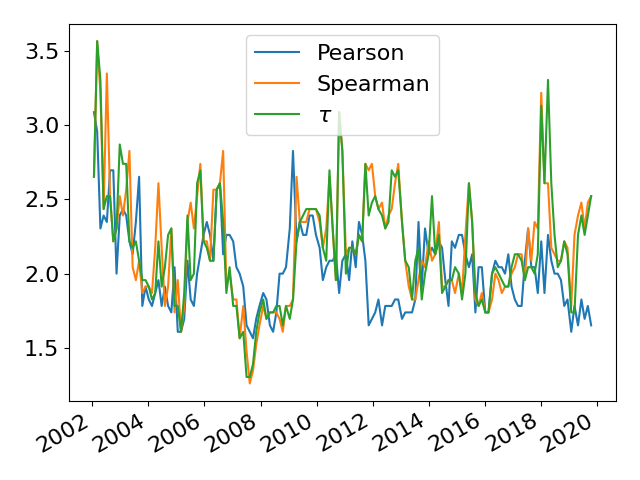}
  \caption{DE}
    \end{subfigure}%
  \caption{Mean Occupation Layer}
  \label{fig:mol}

    \begin{subfigure}{.3\textwidth}
  \includegraphics[width=\textwidth]{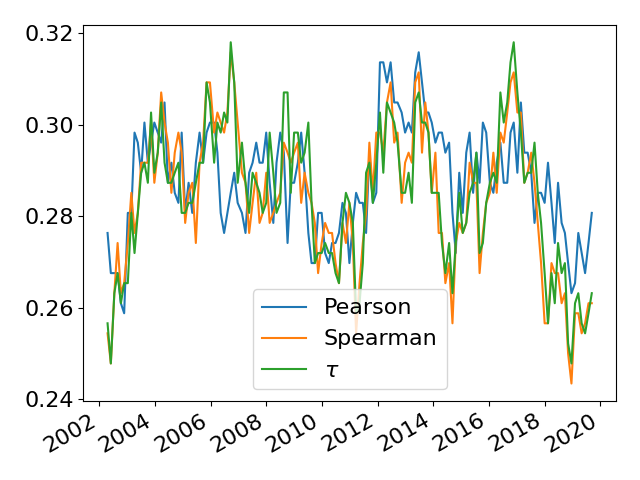}
  \caption{US}
    \end{subfigure}
        \begin{subfigure}{.3\textwidth}
  \includegraphics[width=\textwidth]{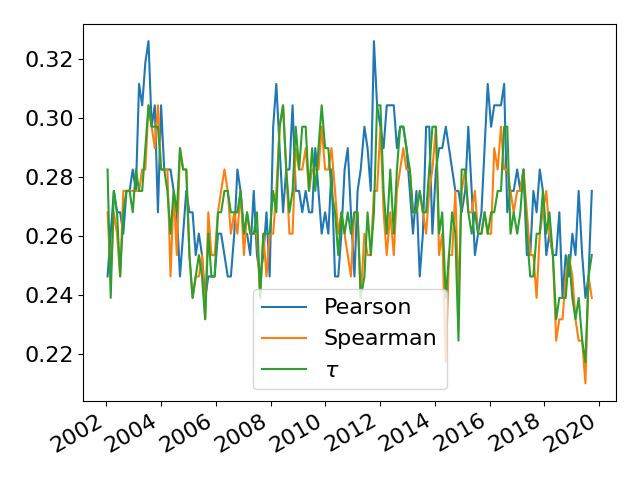}
  \caption{UK}
    \end{subfigure}%
        \begin{subfigure}{.3\textwidth}
  \includegraphics[width=\textwidth]{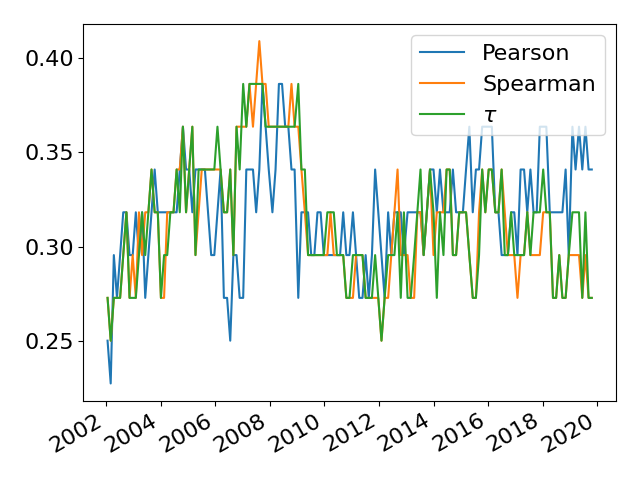}
  \caption{DE}
    \end{subfigure}%
    \caption{Leaf Fraction}
    \label{fig:leaf_fraction}

    \begin{subfigure}{.3\textwidth}
  \includegraphics[width=\textwidth]{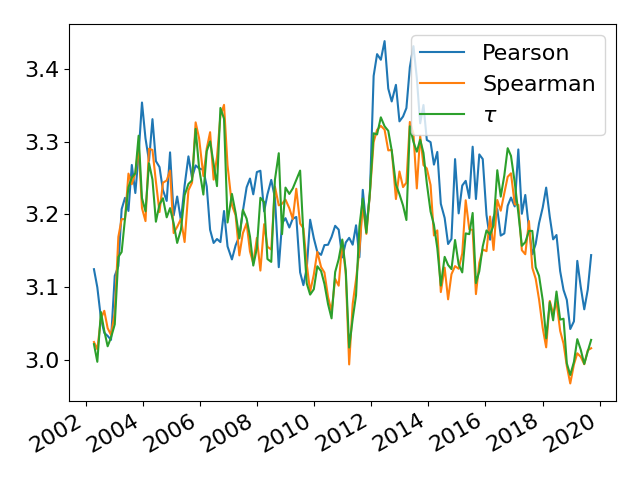}
  \caption{US}
    \end{subfigure}
        \begin{subfigure}{.3\textwidth}
  \includegraphics[width=\textwidth]{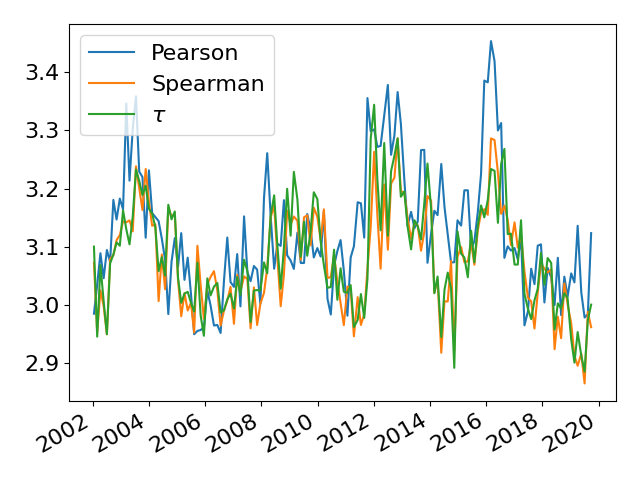}
  \caption{UK}
    \end{subfigure}%
        \begin{subfigure}{.3\textwidth}
  \includegraphics[width=\textwidth]{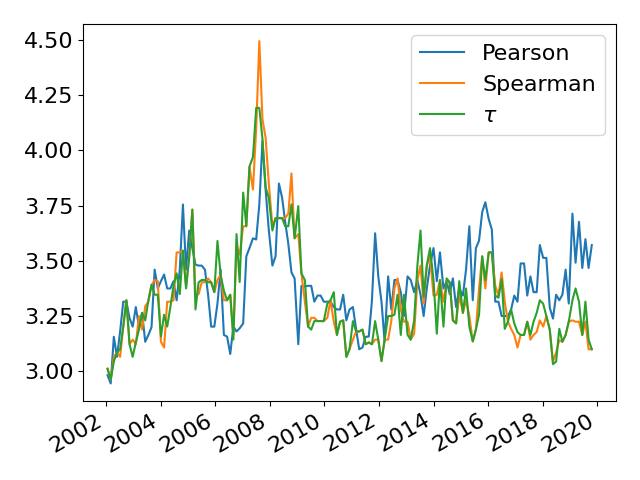}
  \caption{DE}
    \end{subfigure}
  \label{fig:exponent}
  \caption{Exponent}
\end{figure}

Having studied the stability of the trees over time and the importance of various sectors, we now look if the structure of the MSTs differ using some network measures. We use the leaf fraction (fraction of nodes with only 1 edge), exponent when fitting a power law to the degree distribution, the average shortest path length and mean occupation layer (the mean of all shortest paths from each node to the center of the tree). In this case we take the center of the tree as the node with the largest degree. The plots of these measures over time are shown in Figures \ref{fig:av_shortest_path} to \ref{fig:exponent}.

From these we can see that irrelevant of the coefficient, the MSTs have similar structure. All of the trees have a heavy tailed degree distribution, with there being a high number of nodes with only one other edge and a small number of edges with a large degree. The structure of the trees does tend to be dependent on market state for all countries. The average shortest length path is slightly longer for the $\tau$ MSTs than the Pearson or Spearman, but the $\tau$ correlation tends to be slightly smaller for the same value (see Figure \ref{fig:correlation_comparison}) which would explain the longer paths, as they will have a greater distance. 

\subsection{Applications to Portfolio Selection}

As mentioned in the introduction, one of the goals in studying the correlations between financial assets is to assess risk. An obvious application is therefore to construct low risk portfolios using said correlations. MST correlation matrices have been applied for this in previous work \cite{TOLA2008235} with promising results, and in this section we look at how the rank MST correlation matrices compare on this front. To do this, we take the MST filtered correlation matrices and turn them into covariance matrices using
\begin{equation}
    \Sigma = D^{\frac{1}{2}} C D^{\frac{1}{2}}
\end{equation}
where $D$ is a diagonal matrix containing the variances of the assets.

We have found that sometimes the MST correlation matrices are singular, and therefore add a small amount to the diagonal to ensure the resulting correlation matrix is positive definite as follows
\begin{equation}
    \Sigma_* = \alpha \Sigma + (1 - \alpha) tr(\Sigma) I
\end{equation}
where $\alpha=0.9$. This is also applied to the full covariance matrix to assist comparisons. With these resulting covariance matrices, we create minimum risk portfolios by solving the following optimization problem
\begin{equation}
\begin{aligned}
& \underset{\vec{w}}{\text{minimize}}
& &\vec{w}^T \Sigma_* \vec{w} \\
& \text{subject to}
& & \vec{1}^T \vec{w} = 1 \\
& & & w_i > 0 \\
\end{aligned}
\end{equation}
The resulting vector $\vec{w}$ gives us the weight for each asset. 
\begin{figure}
    \centering
    \begin{subfigure}{0.3\textwidth}
      \includegraphics[width=\textwidth]{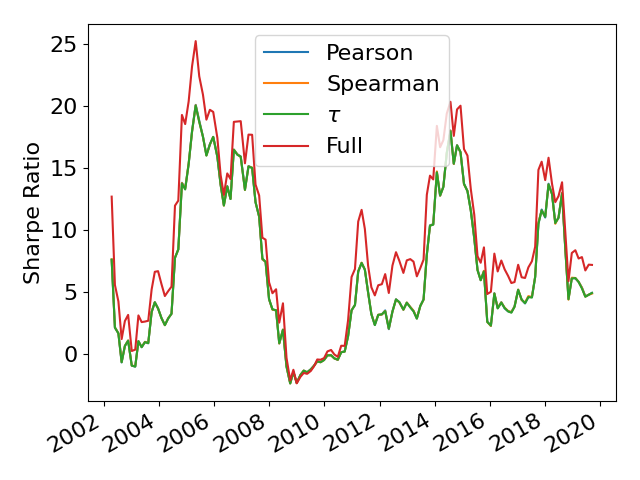}
      \caption{US}
    \end{subfigure}
    \begin{subfigure}{0.3\textwidth}
      \includegraphics[width=\textwidth]{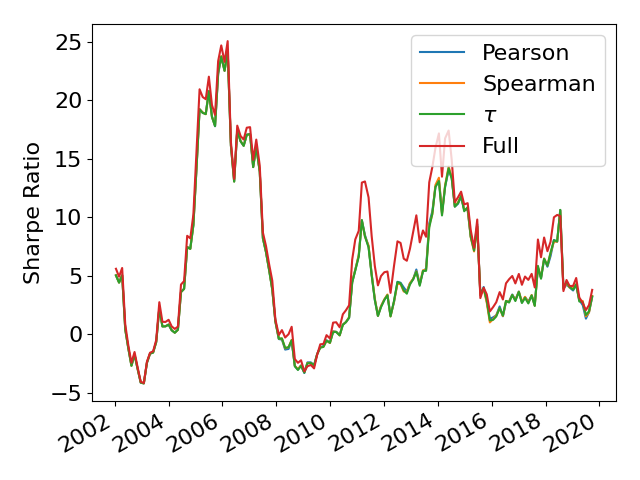}
      \caption{UK}
    \end{subfigure}
    \begin{subfigure}{0.3\textwidth}
      \includegraphics[width=\textwidth]{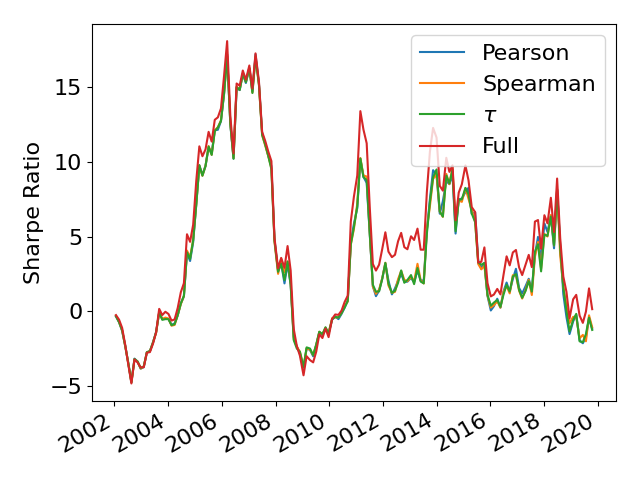}
      \caption{DE}
    \end{subfigure}
    \caption{Out of sample Sharpe ratio of the portfolios. The full covariance matrix generally has a higher Sharpe ratio than the MST covariance matrices, but the difference is not particularly large. For the larger markets this also comes at the cost of a larger portfolio turnover.}
    \label{fig:portfolio_sharpe}
\end{figure}

\begin{figure}
    \centering
    \begin{subfigure}{0.3\textwidth}
      \includegraphics[width=\textwidth]{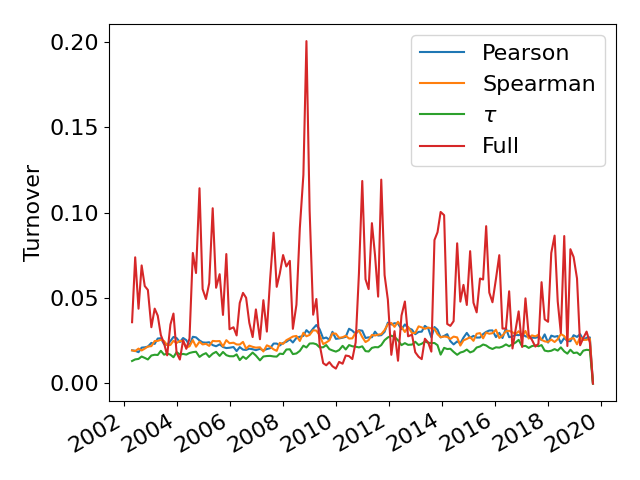}
      \caption{US}
    \end{subfigure}
    \begin{subfigure}{0.3\textwidth}
      \includegraphics[width=\textwidth]{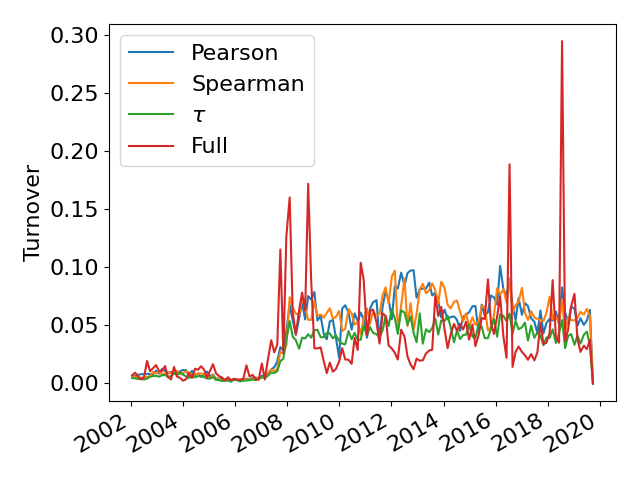}
      \caption{UK}
    \end{subfigure}
    \begin{subfigure}{0.3\textwidth}
      \includegraphics[width=\textwidth]{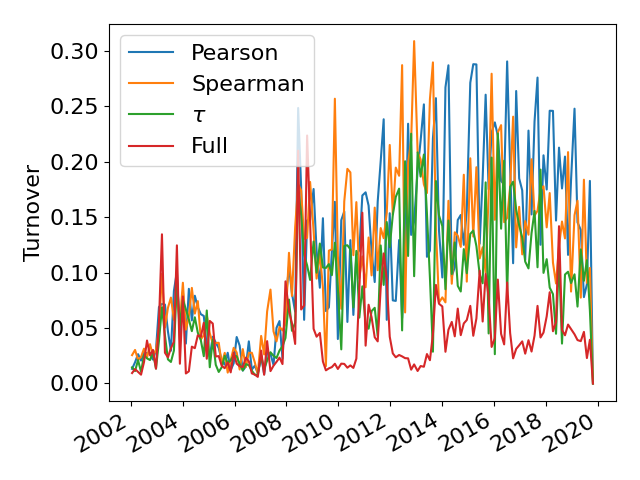}
      \caption{DE}
    \end{subfigure}
    \caption{Turnover of the portfolios constructed using the MST correlation matrices. The MST portfolios tend to have a lower turnover than the full covariance portfolios for the US and UK markets, but not for the German market. Out of the three MSTs, the $\tau$ portfolios have the lowest turnover. }
    \label{fig:portfolio_turnover}
\end{figure}

Firstly we look at the out of sample Sharpe ratios of the resulting portfolios on the following window of data. The results of this are shown in Figure \ref{fig:portfolio_sharpe}. The results for all four portfolios are relatively similar and highly affected by market conditions, but in general the full correlation matrices have a higher Sharpe ratio than the MST filtered ones.
Next we look at the turnover of the portfolios. Since we have found that the rank MSTs tend to be more stable than the Pearson MSTs, we look at how this translates into improving portfolio stability. We measure this by using the $L_1$ norm of the difference between two portfolios adjacent in time
\begin{equation}
    \sum_{i=1}^p |w_{t,i} - w_{t-1, i} |
\end{equation} 
This is shown in Figure \ref{fig:portfolio_turnover}. There is a reduction in mean turnover for the MSTs portfolios for the US and the UK, but not for Germany. This could be due to the smaller size of the German markets, causing $n$ to be much larger than $p$, and therefore the estimation of the full covariance matrix will be much better. For all countries the $\tau$ MSTs have the lowest turnover, followed by the Spearman and then the Pearson MSTs. 
This shows that rank MST covariance matrices may be useful in reducing portfolio turnover when the investor is considering a large number of assets.

\subsection{MST Robustness}

Finally we are interested in comparing the robustness of these correlation coefficients. This can be done using a bootstrap based approach, in a similar manner to Tumminello et al. \cite{tumminello2007spanning} and Musciotto et al. \cite{musciotto20181032}. Here we create 1000 pseudo-datasets using a circular bootstrap. A circular bootstrap is a type of block bootstrap where we select data from a continuous stretch of time, and if the end of the dataset is reached we wrap round and start back at the beginning. This tends to be more appropriate for time series data compared to the classic bootstrap due to look ahead effects and potential autocorrelation. With these pseudo-datasets we calculate the correlations between assets and construct MSTs from these correlation matrices. Once we have this set of MSTs, we can compare the edges present in them. Ideally if there is no noise, the data is purely stationary and the methods robust all these MSTs would be the same. Of course this is not the case in real life. To run the bootstrap we take the first 1008 days of data and create 1000 bootstrapped datasets of 504 days.

Using these bootstrapped datasets we measure the mean and standard deviation of the difference between the full correlation matrices, the MST filtered correlation matrices (i.e. weighted MSTs) and the fraction of difference in edge presence across the MSTs (i.e. unweighted MSTs). To measure the difference between the full and MST filtered correlation matrices we take a similar approach as to section \ref{sec:mst_exploration} and normalise the entries of the correlation matrices to sum to 1 and take the sum of the absolute difference for each entry
\begin{equation}
    \sum_{i=1}^p \sum_{j=1}^p |\frac{C^x_{ij}}{M^x} - \frac{C^y_{ij}}{M^y}|
\end{equation}
where $M^x = \sum_{i=1}^p \sum_{j=1}^p C^x_{ij}$.
The results are shown in Table \ref{tab:correlation_p_value_edge_weight}. If we look at the full correlation matrices, the US and Germany both see a reduction in the mean difference when using rank correlation as opposed to Pearson correlation. For the UK there is a reduction in mean difference for the Spearman MSTs, but not for the $\tau$ MSTs when compared the the Pearson MSTs.
\begin{table}[]
    \centering
    \begin{tabular}{|c|c|c|c|c|c|c|}
    \hline
    Method & \multicolumn{2}{|c|}{MST Weighted} & \multicolumn{2}{|c|}{MST Unweighted} &
    \multicolumn{2}{|c|}{Full} \\
    \cline{2-7}
    & Mean & S.D & Mean & S.D & Mean & S.D \\
    \hline
    \multicolumn{7}{|c|}{US} \\
    \hline
    Pearson & 0.835 & 0.218 & 0.722 & 0.056 & 0.234 & 0.094 \\
    Spearman & 0.830 & 0.209 & 0.721 & 0.053 & 0.175 & 0.074 \\
    $\tau$ & 0.824 & 0.210 & 0.720 & 0.053 & 0.174 & 0.071  \\
    \hline
    \multicolumn{7}{|c|}{UK} \\
    \hline
    Pearson & 0.896 & 0.214 & 0.750 & 0.054 & 0.296 & 0.137 \\
    Spearman & 0.904 & 0.220 & 0.749 & 0.056 & 0.247 & 0.123 \\
    $\tau$ & 0.890 & 0.220 & 0.747 & 0.056 & 0.248 & 0.123  \\
    \hline
    \multicolumn{7}{|c|}{DE} \\
    \hline
    Pearson & 0.732 & 0.249 & 0.690 & 0.064 & 0.226 & 0.101 \\
    Spearman & 0.700 & 0.235 & 0.690 & 0.063 & 0.128 & 0.058 \\
    $\tau$ & 0.665 & 0.231 & 0.684 & 0.062 & 0.121 & 0.048  \\
    \hline
    \end{tabular} 
    \caption{Mean and standard deviation (s.d.) of the difference between full correlation matrices and MSTs constructed from the bootstrapped datasets. For the all of the countries there is a reduction in the mean difference between full correlation matrices when the rank correlation method is used, but there is little reduction for the MST networks, weighted or unweighted.}
    \label{tab:correlation_p_value_edge_weight}
\end{table}
However if we look at the mean difference between the MSTs, the results differ. If we look at the weighted edges there is a slight reduction in difference for the US and Germany, but particularly for the US this is not large. However for the unweighted MSTs there seems to be little to no difference between the MSTs for the US and Germany, and in fact an increase for the UK. From this we would conclude that the MST construction procedure has a larger effect on the robustness of the trees than the correlation coefficient chosen. 

\section{Conclusion}

In this paper we have used the Pearson, Spearman and Kendall's $\tau$ correlation coefficients to infer correlation matrices from stock returns from three countries (the US, UK and Germany), constructed minimum spanning trees from these matrices and compared the robustness and evolution of the trees over time. 

Looking at the evolution of the trees over the dataset we have found the MSTs constructed using the rank correlations (Spearman and Kendall's $\tau$) change less than Pearson MSTs (notably during times of market stress) and have edges that are maintained for a longer time period over the dataset. Despite this, the trees tend to have a similar topology, irrelevant of coefficient and this topology tends to vary in a similar way over time for all three methods. Perhaps unsurprisingly, the structure of the rank MSTs is very similar, while they both differ more from the Pearson MSTs.

In general all of the MSTs tend to show broad agreement on which sectors are regarded as important over the entire dataset, but there can be significant disagreements at particular points in time. The rank MSTs show more agreement with each other than either with the Pearson MSTs. The agreement using degree centrality is higher than using betweenness centrality, indicating that companies tend to be found in different places on the trees in the Pearson MSTs compared to the rank ones.

We then attempt to connect departures from univariate Gaussianity for individual companies to changes in their expression in the MSTs and full correlation matrices. These deviations are correlated with changes in the expression of a company in the full correlation matrices, but not in the MSTs. Furthermore we then use a quantile normalisation method to enforce univariate Gaussianity on each company, and then run the analysis again. From this we find that there is a reduction in difference between the rank and Pearson MSTs, but there is still a much larger difference between them than between the Spearman and $\tau$ MSTs, indicating that there could be non-linear relationships involved too. 

These MSTs can also be applied for use in portfolio selection. We construct minimum risk portfolios from the MST correlation matrices and compare the resulting portfolios to those produced using the full covariance matrix. The portfolios constructed from the MST correlation matrices tend to have a lower turnover compared to those constructed using the full matrices for the larger markets, but tend to have a slightly lower Sharpe ratio. In particular for the portfolios constructed the MSTs, the $\tau$ portfolios have the lowest turnover, while their Sharpe ratio is indistinguishable from those constructed from the Pearson MST correlation matrices. 

Finally we use a bootstrap to test the consistency of the correlation matrices inferred and the edges selected in the MSTs. We find that the full correlation matrices constructed using rank correlations mostly have a smaller difference than the full Pearson correlation matrices, but there is relatively little difference in the MSTs, indicating the MST construction procedure has a larger influence on this than the correlation coefficient chosen.

Overall it may be worth constructing MSTs using different correlation coefficients for a given dataset to give a different picture, but the MST construction procedure has the greatest influence on the results.
Generally the Spearman and Kendall's $\tau$ correlation coefficients tend to give similar results, indicating that if computational resources are constrained then calculating the Spearman correlation is sufficient. Future work could proceed in several directions. A comparison of mutual information MSTs to these correlation MSTs to see how they differ could be interesting, or exploring different filtration models, for instance the Planar Maximally Filtered Graph. Alternatively these comparisons could be performed with returns data from other countries or assets, perhaps from data that is highly correlated and volatile, for instance for returns from cryptocurrencies or developing nations.  

\section*{Funding}
 TM Acknowledges PhD studentship funding from the School of Electronics and Computer Science, University of Southampton; 



\bibliographystyle{elsarticle-num} 
\bibliography{used}


\end{document}